%% file: main.tex
\documentclass[twocolumn,apj]{openjournal}
\usepackage{orcidlink}
\usepackage{graphicx}
\usepackage{amsmath}
\usepackage{amssymb}
\usepackage{natbib}
\usepackage{booktabs}
\usepackage{multirow}
\usepackage{longtable}
\hypersetup{colorlinks=true,linkcolor=blue,citecolor=blue,urlcolor=blue}
\usepackage{xcolor}
\usepackage{url}

\newif\ifdraft
\drafttrue
\ifdraft
  \newcommand{\todo}[1]{\textcolor{red}{\textbf{[TODO: #1]}}}
  \newcommand{\note}[2]{\textcolor{blue}{\textbf{[#1: #2]}}}
\else
  \newcommand{\todo}[1]{}
  \newcommand{\note}[2]{}
\fi

\newcommand{\atlas}{ATLAS}
\newcommand{\ofilt}{\ensuremath{o}}
\newcommand{\cfilt}{\ensuremath{c}}
\newcommand{\como}{\ensuremath{c-o}}

\providecommand{\degr}{\ensuremath{^{\circ}}}
\newcommand{\ema}{EMA}
\newcommand{\mbr}{MBR Explorer}

\newcommand{\justitia}{(269)~Justitia}
\newcommand{\chimaera}{(623)~Chimaera}
\newcommand{\westerwald}{(10253)~Westerwald}
\newcommand{\rockox}{(13294)~Rockox}
\newcommand{\ousha}{(23871)~Ousha}
\newcommand{\moza}{(59980)~Moza}
\newcommand{\ghaf}{(88055)~Ghaf}

\begin{document}

\title{Long-baseline ATLAS photometry of the seven main-belt asteroid targets
       of the Emirates Mission to the Asteroid Belt}


\author{Nicolas Erasmus\,\orcidlink{0000-0002-9986-3898}}
\affiliation{South African Astronomical Observatory, Cape Town, 7925, South Africa}
\affiliation{Department of Physics, Stellenbosch University, Stellenbosch, 7602, South Africa}

\author{Josef {\v D}urech\,\orcidlink{0000-0003-4914-3646}}
\affiliation{Astronomical Institute, Faculty of Mathematics and Physics,
Charles University, V Hole\v{s}ovi\v{c}k\'ach 2, 180 00 Prague, Czech Republic}

\author{David E. Trilling\,\orcidlink{0000-0003-4580-3790}}{}
\affiliation{Department of Astronomy and Planetary Science, PO Box 6010, Northern Arizona University, Flagstaff, AZ 86011, USA}

\author{Larry Denneau\,\orcidlink{0000-0002-7034-148X}}
\affiliation{Institute for Astronomy, University of Hawaii, 2680 Woodlawn Drive,
Honolulu, HI 96822, USA}

\author{John L. Tonry\,\orcidlink{0000-0003-2858-9657}}
\affiliation{Institute for Astronomy, University of Hawaii, 2680 Woodlawn Drive,
Honolulu, HI 96822, USA}

\author{Henry Weiland}
\affiliation{Institute for Astronomy, University of Hawaii, 2680 Woodlawn Drive,
Honolulu, HI 96822, USA}

\author{Ken W. Smith\,\orcidlink{0000-0001-9535-3199}}
\affiliation{Department of Physics, University of Oxford, Denys Wilkinson
Building, Keble Road, Oxford OX1 3RH, UK}
\affiliation{Astrophysics Research Centre, School of Mathematics and Physics,
Queen's University Belfast, BT7 1NN, UK}

\author{Javier Licandro\,\orcidlink{0000-0002-9214-337X}}
\affiliation{Instituto de Astrofísica de Canarias (IAC), C/ Vía Láctea, s/n, E-38205, La Laguna, Spain}
\affiliation{Departamento de Astrofísica, Universidad de La Laguna (ULL), E-38206 La Laguna, Canarias, Spain}

\author{Miguel R. Alarcon\,\orcidlink{0000-0002-8134-2592}}
\affiliation{Light Bridges, Observatorio Astronómico del Teide. Carretera del Observatorio del Teide, s/n, Güímar, Canarias, Spain}
\affiliation{Instituto de Astrofísica de Canarias (IAC), C/ Vía Láctea, s/n, E-38205, La Laguna, Spain}
\affiliation{Departamento de Astrofísica, Universidad de La Laguna (ULL), E-38206 La Laguna, Canarias, Spain}

\begin{abstract}
The Emirates Mission to the Asteroid Belt (EMA) will use the \mbr{} spacecraft
to fly past six main-belt asteroids --- \westerwald{}, \chimaera{}, \rockox{},
\ghaf{}, \ousha{} and \moza{} --- before rendezvousing with, and deploying a
lander onto, the extremely red asteroid \justitia{}. We present a homogeneous
analysis of roughly a decade of serendipitous sparse photometry of all seven
targets obtained with the Asteroid Terrestrial-impact Last Alert System
(ATLAS), retrieved through the ATLAS forced photometry service. For each object
we derive absolute magnitudes and phase slope parameters in both the ATLAS
\cfilt{} and \ofilt{} filters, the \como{} colour on a common phase slope, a
rotation period from a Lomb--Scargle analysis of the combined, phase-corrected
light curve, and a light-curve amplitude that sets a lower limit on the body's
elongation. We recover rotation periods spanning $3.2$ to $33$~h. Five are
secure --- $3.243$~h for \rockox{}, $33.1$~h for \justitia{}, $3.64$~h for
\westerwald{}, $14.6$~h for \chimaera{} and $4.36$~h for \moza{} --- each agreeing with independent
determinations; for \ousha{} our periodograms are affected by
diurnal aliasing and we defer to the published period of $8.35$~h, while the period of \ghaf{} remains unresolved. We further derive a unique convex shape and
spin-state model for \rockox{} by lightcurve inversion and shape models with two possible spin-pole solutions for \chimaera{}, \moza{} and \justitia{}. No shape models could be determined for \ghaf{}, \ousha{} or \westerwald{} but  we can constrain spin-axis orientation to retrograde for \westerwald{}. Combining the
\como{} colour with an albedo class inferred from the phase slope, we classify
the sample as one dark C-type (\chimaera{}), three probable S-types
(\westerwald{}, \ghaf{}, \moza{}), two X-complex objects (\rockox{}, plausibly
M-type, and \ousha{}, possibly E-type), and the exceptionally red L/D-type
\justitia{}, whose $(c-o) = 0.48$ is redder than typical Jupiter Trojans. This
distribution only partly matches the mission's family-based expectation of a
predominantly primitive target set, so we present the classifications as
testable pre-encounter predictions. We caution that our classifications of
\ghaf{} and \ousha{} are the least secure in the sample: both disagree with
previously published classifications, and both are also the targets whose
rotation periods are unresolved or alias-affected. We therefore highlight these
two objects as warranting additional follow-up ahead of encounter.
These results provide
a homogeneous physical characterisation of the EMA target set ahead of
encounter.\\\\
\end{abstract}


\input{sections/01_introduction}
\input{sections/02_atlas}
\input{sections/03_methods}
\input{sections/04_results}
\input{sections/06_discussion}
\input{sections/07_conclusions}
\input{sections/08_acknowledgements}

\bibliographystyle{aasjournal}
\bibliography{refs}

\appendix
\input{sections/A1_figures}

\end{document}

%% file: sections/01_introduction.tex
\section{Introduction}
\label{sec:intro}

The Emirates Mission to the Asteroid Belt (EMA) is a main-belt asteroid tour
led by the UAE Space Agency\footnote{\url{https://space.gov.ae/en/projects-and-initiatives/space-exploration/emirates-mission-to-the-asteroid-belt}}
and announced in May 2023. The mission spacecraft is scheduled for launch in 2028 from the
Tanegashima Space Center on an H3 launch vehicle. Using a combination of Venus,
Earth and Mars gravity assists together with solar electric propulsion, the
spacecraft will travel approximately $5\times10^{9}$~km over a nominal
seven-year cruise \citep{parker2024}. \ema{} is the first multiple-asteroid tour
and landing mission directed at the main belt, and is in several respects the
main-belt analogue of NASA's \emph{Lucy} mission to the Jupiter Trojans
\citep{levison2021}.

The primary objective of \ema{} is to rendezvous with, and conduct orbital
operations at, a single main-belt asteroid, \justitia{}, culminating in the
deployment of a small lander to its surface; the secondary objective is to
explore six further main-belt asteroids via high-speed flybys en route
\citep{parker2024}. The nominal encounter sequence is summarised in
Table~\ref{tab:targets}. The seven targets span a wide range of size, dynamical
association and inferred composition, and several are thought to be primitive,
potentially water-rich C-complex bodies \citep{alsaeed2025}. By sampling this
diverse set the mission aims to fill gaps in our understanding of the variety of
objects in the main belt and, through them, the origin and evolution of the
Solar System.

The rendezvous target, \justitia{}, is the scientifically most distinctive of
the seven. A $\sim$57~km object \citep{buie2025}, it was identified together with (203)~Pompeja
as one of the two reddest bodies known in the main asteroid belt
\citep{hasegawa2021}: its spectral slope exceeds that of D-type asteroids and
more closely resembles the very red Centaurs and trans-Neptunian objects,
suggesting a surface rich in complex organic material and, possibly, an origin
in the outer Solar System followed by inward migration during the early
dynamical evolution of the planetary system \citep{hasegawa2021,marciniak2025}.
Recent thermophysical modelling give a diameter of
$\sim$55--60~km, a low geometric albedo of $\sim$0.06--0.08, and a rotation
period of $33.13$~h \citep{marciniak2025}. \justitia{} is therefore a natural
high-priority target for \emph{in situ} investigation.

Ground-based characterisation of flyby targets ahead of encounter is essential
for encounter planning: rotation periods and pole orientations constrain the
sub-spacecraft geometry and hence which hemisphere will be illuminated and
imaged; absolute magnitudes and phase functions constrain sizes and albedos
and therefore exposure planning; and broad-band colours provide a first-order
taxonomic classification that informs spectral observation sequencing. For the
smaller \ema{} targets in particular, the existing physical characterisation is
sparse.

In this work we exploit approximately a decade of serendipitous, sparse
photometry of all seven \ema{} targets obtained by the Asteroid
Terrestrial-impact Last Alert System (\atlas{}), retrieved through the public
\atlas{} forced photometry service. We derive homogeneous absolute magnitudes,
phase slope parameters and \como{} colours in the two \atlas{} filters, and
search for rotation periods in the combined, phase-corrected light curves. The
paper is organised as follows. Section~\ref{sec:atlas} describes the \atlas{}
survey, the forced photometry service, and the value of long-baseline sparse
photometry for main-belt science. Section~\ref{sec:methods} sets out our
analysis pipeline, including the convex lightcurve inversion used to derive
shape and spin-state models.
Section~\ref{sec:results} presents the results for each target in turn ---
photometric parameters, rotation period and shape model --- together with a
comparison to published values. We discuss the implications in
Section~\ref{sec:discussion} and conclude in Section~\ref{sec:conclusions}.

\begin{table}
\centering
\caption{The seven target asteroids of the Emirates Mission to the Asteroid
Belt, in encounter order, with encounter dates from \citet{parker2024}.}
\label{tab:targets}
\begin{tabular}{lll}
\toprule
Target & Encounter type & Encounter date$^{\ast}$ \\
\midrule
\westerwald{} & Flyby            & 2030 Feb \\
\chimaera{}   & Flyby            & 2030 Jun \\
\rockox{}     & Flyby            & 2031 Jan \\
\ghaf{}       & Flyby            & 2032 Jul \\
\ousha{}      & Flyby            & 2032 Dec \\
\moza{}       & Flyby            & 2033 Aug \\
\justitia{}   & Rendezvous + lander & 2034 Oct \\
\bottomrule
\end{tabular}

\vspace{2pt}
\begin{minipage}{\linewidth}
\footnotesize $^{\ast}$ Dates available at the time of writing; these remain
subject to change as the mission design matures and the launch date is
finalised.
\end{minipage}
\end{table}

%% file: sections/02_atlas.tex
\section{The ATLAS survey as a resource for main-belt science}
\label{sec:atlas}

\subsection{The ATLAS telescope network}
\label{sec:atlas_network}

The Asteroid Terrestrial-impact Last Alert System \citep[\atlas{};][]{tonry2018}
is a NASA-funded near-Earth object early warning survey operated by the
University of Hawai\'{}i Institute for Astronomy. The system began operations in
2015 with a single 0.5~m telescope on Haleakal\=a (MPC code T05), joined in 2017
by a second identical unit on Mauna Loa (T08). The network was extended into
the southern hemisphere in 2022 with two further identical 0.5~m units, at
Sutherland in South Africa (M22), operated by the South African Astronomical
Observatory, and at Rio Hurtado / El~Sauce in Chile (W68). These four units
share a common optical and detector design and together survey the accessible
sky every $\sim$1--2 nights.

Each unit images in two broad, non-standard filters: cyan
($c$, $\sim$420--650~nm) and orange ($o$, $\sim$560--820~nm), with the choice
driven principally by lunation --- the cyan filter is used around full moon,
while the orange filter is used for the rest of the lunation.
As a result, roughly three times as many $o$-band as $c$-band measurements are
typically available for a given object.
Images are obtained in a
standard pattern of four 30~s exposures spread over roughly an hour, a cadence
designed for the linkage of moving objects, and reach a typical limiting
magnitude of $o\simeq19.5$. The pixel scale is $\sim1.9''$~pixel$^{-1}$.
Astrometric and photometric calibration is performed against the \atlas{}
All-Sky Stellar Reference Catalog \citep[RefCat2;][]{tonry2018b}, and the
resulting photometry is on the AB system.

A fifth \atlas{} unit, \atlas{}-Teide, was subsequently commissioned at the
Teide Observatory in Tenerife (R17) \citep{licandro2023}. This unit differs
substantially from the other four in design --- it comprises four modules of
four commercial telescopes each, together equivalent to a $\sim$56~cm aperture
--- and observes in a single broad \emph{wide} ($w$) filter rather than the
$c/o$ pair. Because the \atlas{}-Teide unit observes in a single $w$ filter, it
cannot contribute to the $c-o$ colour analysis on which our taxonomic inference
relies. It also adds only a single apparition of data, against the several
apparitions already provided by the four original units over the survey
baseline, so its contribution to the phase curves and rotation search would be
marginal. We therefore exclude all \atlas{}-Teide data from this study. As the
Teide baseline lengthens, however, its additional apparitions should be
incorporated in future work, particularly to improve rotation period
determination. This will be especially important for targets that only have a single short apparition like near-Earth asteroids.

\subsection{The ATLAS forced photometry service}
\label{sec:atlas_fps}

Although \atlas{} is optimised for the detection of moving objects and
transients, all calibrated exposures are retained on disk, which makes it
possible to recover photometry at any sky position and epoch after the fact.
The \atlas{} forced photometry
service\footnote{\url{https://fallingstar-data.com/forcedphot/}}
\citep{shingles2021} provides public access to this capability. Provided a fixed
RA and Dec position or for minor bodies a designation, the service performs PSF-fitting photometry at the specified location on every
archival exposure covering it, using the \texttt{tphot} routine
\citep{sonnett2013} applied to either the reduced target image or the
difference image. For this study we used the difference images to minimize contamination by background sources. The returned light curve includes serval columns but relevant to our data analysis the measured flux, the corresponding magnitude and
uncertainty, the filter, the exposure MJD, the
$5\sigma$ limiting magnitude of the image, and the contributing site. After data extraction, we perform some careful quality filtering as described in Section~\ref{sec:methods_scrub}.

\subsection{Sparse, long-baseline photometry of main-belt asteroids}
\label{sec:atlas_sparse}

The \atlas{} archive now spans roughly a decade. For a typical main-belt
asteroid this corresponds to several complete apparitions, sampled at a cadence
of order one to two nights (mostly 4 data-points per night) whenever the object is above the survey limit. The
resulting light curves are \emph{sparse} --- individual nights contribute only
a handful of points, and consecutive epochs are separated by intervals far
longer than a typical rotation period --- but they are also long, dense in
total, well calibrated, and obtained over a wide range of phase angles and
viewing geometries. This combination is precisely what is required to
simultaneously constrain phase functions, colours and rotation periods, and it
is not readily obtainable from targeted observing campaigns.

A number of studies have demonstrated the power of this approach on large
asteroid samples. \citet{erasmus2020} used \atlas{} dual-band $c-o$ photometry
to investigate taxonomic diversity within asteroid families.
\citet{mahlke2021} derived phase curves for a large sample of asteroids from
\atlas{} dual-band photometry, and \citet{robinson2024} extended phase curve
determinations to main-belt and Trojan asteroids using the \atlas{} survey.
\citet{mcneill2021} compared the physical properties of the L4 and L5 Trojan
populations using \atlas{} data, and \citet{erasmus2021} combined \atlas{} and
ZTF photometry to identify super-slow rotators. Sparse \atlas{} photometry has
also proven sufficient for full lightcurve inversion: \citet{durech2020}
reconstructed convex shape models and spin states for many hundreds of
asteroids from \atlas{} photometry alone, and \citet{durech2023} extended this
using bootstrap convex inversion to derive rotation periods at scale.

What is notable about this body of work is that it is almost exclusively
\emph{population-level}: \atlas{} data are used to characterise the statistical
properties of large samples, rather than to build a detailed picture of any
individual object. In this paper we invert that emphasis, applying the same
survey resource to a small, pre-selected set of seven objects of specific
mission relevance, and extracting from each the fullest physical
characterisation the data support.

%% file: sections/03_methods.tex
\section{Data reduction and analysis}
\label{sec:methods}

All seven targets were processed through an identical pipeline, so that the
derived parameters can be compared between objects without survey- or
method-dependent systematics. The pipeline is implemented in Python and makes use
of \texttt{astropy} \citep{astropy2022}, \texttt{numpy} \citep{harris2020},
\texttt{scipy} \citep{virtanen2020} and \texttt{matplotlib}
\citep{hunter2007}.

\subsection{Data retrieval and quality filtering}
\label{sec:methods_scrub}

For each target, forced photometry was requested from the \atlas{} service
(Section~\ref{sec:atlas_fps}) and per-epoch observing geometry --- phase angle $\alpha$, heliocentric distance
$R$ and geocentric distance $\Delta$ --- was queried from JPL Horizons via \texttt{astropy} and
cached locally.

Raw forced photometry contains a substantial fraction of unusable measurements,
arising from non-detections, exposures affected by cloud or bright moonlight,
and cases where the ephemeris position falls on or near a bright background source that isn't completely removed in the difference image.
We applied the following cuts, using the default parameters of the pipeline's
cleaning routine:
\begin{enumerate}
  \item A quality cut requiring the image $5\sigma$ limiting magnitude to be
        fainter than 18.0, the detected flux to exceed 30~$\mu$Jy, and the
        photometric uncertainty to satisfy $\sigma_m \leq 0.3$~mag.
  \item An initial crude outlier rejection on apparent magnitude: measurements more than
        1.5~mag from the per-object median apparent magnitude were discarded.
  \item Only $c$- and $o$-band measurements were retained; \atlas{}-Teide
        $w$-band data are not processed (Section~\ref{sec:atlas_network}).

\end{enumerate}
Observation times were corrected for light travel time to give
$\mathrm{MJD}_{\mathrm{corr}} = \mathrm{MJD}_{\mathrm{obs}} - \Delta/c$, and
for the mid-exposure offset. The surviving number of $c$ and $o$
data points per target as well as the MJD and phase angle range are summarised in Table~\ref{tab:obs_summary}

\begin{table}
\centering
\caption{Summary of the \atlas{} forced photometry used in this work, after
quality filtering and outlier rejection. $N_o$ and $N_c$ are the numbers of
surviving $o$- and $c$-band measurements; the baseline and phase-angle range
are for the combined data.}
\label{tab:obs_summary}
\setlength{\tabcolsep}{4pt}
\begin{tabular}{lrrcc}
\toprule
Target & $N_o$ & $N_c$ & Baseline (MJD) & $\alpha$ (\degr) \\
\midrule
\westerwald{} & 1261 & 385 & 57313--61034 & 0.0--30.4 \\
\chimaera{}   & 1938 & 510 & 57430--61128 & 5.1--26.7 \\
\rockox{}     & 1756 & 718 & 57228--61048 & 1.4--29.9 \\
\ghaf{}       &  839 & 354 & 57465--61245 & 0.2--23.6 \\
\ousha{}      &  707 & 271 & 57602--61134 & 0.3--31.2 \\
\moza{}       & 2158 & 556 & 57227--61228 & 0.2--19.7 \\
\justitia{}   & 2501 & 663 & 57314--61247 & 0.7--29.0 \\
\midrule
Total         & 11160 & 3457 & & \\
\bottomrule
\end{tabular}
\end{table}

\subsection{Reduced magnitudes and phase curves}
\label{sec:methods_phase}

Apparent magnitudes were converted to reduced magnitudes,
\begin{equation}
  m_{\mathrm{red}} = m - 5\log_{10}(R\,\Delta),
  \label{eq:reduced}
\end{equation}
removing the distance dependence and leaving the phase-angle dependence
explicit. For each filter independently we fitted the $H,G$ phase function
\citep{bowell1989},
\begin{equation}
  m_{\mathrm{red}}(\alpha) = H - 2.5\log_{10}
    \left[(1-G)\,\Phi_1(\alpha) + G\,\Phi_2(\alpha)\right],
  \label{eq:hg}
\end{equation}
with the basis functions evaluated in the form given by \citet{dymock2007},
\begin{equation}
  \Phi_i(\alpha) = \exp\left[-A_i \tan^{B_i}(\alpha/2)\right],
\end{equation}
where $(A_1,B_1) = (3.332, 0.631)$ and $(A_2,B_2) = (1.862, 1.218)$.

Fits were performed with a bounded Levenberg--Marquardt least-squares
minimisation. Because the quality-filter cut on \emph{apparent} magnitude
does not remove points that are outliers in \emph{reduced} magnitude --- for
example, cosmic rays or centroid errors near a background source at a similar flux, which
typically carry small reported uncertainties and so survive per-point quality
cuts --- we applied an iterative rejection on the fit residuals. At each
iteration the phase function was fitted, residuals computed, and points more
than $3\sigma$ from the residual median rejected, where $\sigma$ is estimated
robustly from the median absolute deviation.
Previously rejected points are re-tested at every iteration and may return.
The procedure iterates until the mask is stable or five iterations are reached.
Rejected points were typically only 10--20 per object and remain visible as
obvious outliers in the phase-curve panels of the per-target diagnostic plots
(Figures~\ref{fig:westerwald}--\ref{fig:justitia}, Appendix~\ref{app:figures}).
\subsection{Absolute magnitudes and colours}
\label{sec:methods_colour}

Because the $c$ and $o$ observations are not simultaneous, the \como{} colour
cannot be measured directly on a per-epoch basis; rotational variability would
dominate any pairwise difference. We instead derive the colour from the
per-filter absolute magnitude distributions, on a common phase system.

The two filters are first fitted independently with
Equation~\ref{eq:hg} to obtain $(H_o, G_o)$ and $(H_c, G_c)$, including the
iterative residual rejection described in Section~\ref{sec:methods_phase}. The
per-filter slopes are then averaged to define a single shared slope,
\begin{equation}
  G_{\mathrm{avg}} = \tfrac{1}{2}(G_o + G_c),
  \label{eq:gavg}
\end{equation}
and each filter's absolute magnitude is re-derived by fitting only $H$ with the
slope held fixed at $G_{\mathrm{avg}}$ (applied to the points that survived the
per-filter outlier rejection). Placing both filters on the same phase slope
ensures that the derived absolute magnitudes, and hence the colour, are not
biased by the difference between the independently fitted $G_o$ and $G_c$ ---
which can be substantial when one filter has sparse or geometrically restricted
phase-angle coverage.

The colour is then the difference of the per-filter median absolute magnitudes,
\begin{equation}
  (c-o) = \mathrm{median}(H_{c,\mathrm{abs}}) - \mathrm{median}(H_{o,\mathrm{abs}}),
  \label{eq:colour}
\end{equation}
which is robust to the residual rotational scatter provided both filters sample
the rotational phase distribution fairly --- a reasonable assumption given the
number of epochs and the long baseline. Note that the above procedure means that \como{} colours reported in this study are \emph{not} the fitted $H_c-H_o$. The uncertainty is estimated by a
Monte-Carlo procedure: we repeatedly draw a random subsample (a fraction 0.5 of
each filter's surviving points, without replacement), take the median of each
subsample, difference the two medians, and adopt the standard deviation of these
200 realisations as the colour uncertainty (random seed 42).

\subsection{Rotation periods}
\label{sec:methods_period}

Rotation periods were searched for in the phase-corrected absolute magnitude
light curve, formed by subtracting the best-fit phase function of
Equation~\ref{eq:hg} from the reduced magnitudes in each filter and combining
the two filters onto a common scale using the colour derived in
Section~\ref{sec:methods_colour}. This removes the slow geometric trends and
leaves the rotational signal.

We used the Lomb--Scargle periodogram
\citep{lomb1976,scargle1982} as implemented in
\texttt{astropy.timeseries.LombScargle} \citep{vanderplas2018}, weighting each
point by its photometric uncertainty. The search was carried out over
light-curve periods from $P_{\min}$ = 0.1 to $P_{\max}$ = 200~h. Because a
typical asteroid light curve is dominated by the second harmonic --- two maxima
and two minima per rotation for a triaxial body --- the rotation period is
taken to be twice the recovered light-curve period,
$P_{\mathrm{rot}} = 2P_{\mathrm{LC}}$, and both are quoted in the periodogram
panels of Figures~\ref{fig:westerwald}--\ref{fig:justitia}, Appendix~\ref{app:figures}. Period uncertainties were
estimated from the width of the periodogram peak.

Sparse ground-based survey data are strongly susceptible to aliasing at periods
commensurate with a sidereal day, where the diurnal sampling window produces
spurious periodogram power. We therefore masked out narrow bands ($\pm$2\%)
around a set of alias periods --- 4, 6, 8, 12, 16, 24, 32, 48 and 96~h --- so
that no peak falling within them could be selected as the best period; these
masked regions are shaded grey and labelled in the periodogram panels
(Figures~\ref{fig:westerwald}--\ref{fig:justitia}, Appendix~\ref{app:figures}).
Among the surviving peaks, the strongest was adopted as the light-curve period,
and the three highest-ranked peaks (subject to a minimum-separation criterion)
were inspected for the characteristic $1{:}2$ and $2{:}1$ ratios that signal an
ambiguity between the true rotation period and its harmonics. Folded light
curves at the adopted period are shown in the final panel of each figure.

\subsection{Light-curve amplitude}
\label{sec:methods_amplitude}

The peak-to-peak light-curve amplitude was measured from the folded, combined
light curve rather than from the raw extrema, which are dominated by
photometric outliers. We computed a boxcar running average over the folded
light curve, using a sliding window of width 10\% of a single
rotational cycle that wraps around the fold boundary so that the smoothed curve
is continuous in phase. The amplitude is then the difference between the
maximum and minimum of this running-average curve, and is quoted in the folded
light-curve panel of each figure. This provides a robust, if conservative,
lower limit on the true amplitude: smoothing suppresses the sharpest extrema,
so the tabulated amplitudes should be read as lower bounds, particularly for
the low-amplitude and noisier targets.


\subsection{Convex shape models from ATLAS photometry}
\label{sec:methods_shapes}

In addition to the phase-curve, colour and periodogram analysis above, we derive
convex shape and spin-state models for the targets by lightcurve inversion,
using the same \atlas{} quality-filter forced-photometry data as explained in Section~\ref{sec:methods_scrub}. 
We use the light curve inversion method of \cite{kaasalainen2001a, kaasalainen2001b} specifically adopted for sparse photometry \citep{kaasalainen2004}. The method used the least-squares minimization to converge to the local minimum in the parameter space, where the parameters of the optimization are the sidereal rotation period $P$, the direction of the spin axis in the ecliptic coordinates $(\lambda, \beta)$, three parameters decribing the phase curve, the $c - o$ colour, and the coefficients of spherical harmonics describing the convex shape model. To find the global minimum, we densely scanned the interval 2--1000~h for the rotation period and ten initial pole directions for each trial period. The same approach has already been applied to ATLAS photometry by \cite{durech2020}, which also describes the details of the method.

If the number of photometric measurements is large and measurement errors are small relative to the light-curve amplitude, the method provides a unique solution for the spin state and the corresponding shape model that fits the data significantly better than all other solutions. In practice, there are often two possible models with the same rotation period but different spin-axis directions. This stems from the symmetry of the problem when the asteroid has a low inclination and viewing and illumination geometry is limited to the plane of ecliptic \citep[ambiguity theorem, ][]{kaasalainen2006}. If the number of data points is small or the amplitude of the signal is lost in noise, the method does not provide a reliable model because there are many local minima in the parameter space that provide practically the same fit to the data. This was the case for asteroids \ghaf{} and \ousha{} as described in the next section.

To estimate uncertainties in the derived spin parameters, we created one thousand bootstrap data sets for each asteroid and repeated the inversion, converging each time to slightly different spin parameters. The bootstrap data were created by randomly resampling photometric measurements in $c$ and $o$ filters separately. We used the standard deviation of the spin parameters and $c - o$ colour as the estimate of the $1\sigma$ error.

%% file: sections/04_results.tex
\section{Results}
\label{sec:results}

We present the results for each target in turn, in order of the \mbr{}
encounter sequence (Table~\ref{tab:targets}): \westerwald{}, \chimaera{},
\rockox{}, \ghaf{}, \ousha{}, \moza{}, and finally the rendezvous target
\justitia{}. For every object a six-panel diagnostic plot is given in
Appendix~\ref{app:figures} (Figures~\ref{fig:westerwald}--\ref{fig:justitia}).
The panels follow the order of the analysis: (top left) apparent magnitude
against date, separated by filter; (middle left) reduced magnitude against
phase angle with the fitted $H,G$ phase functions overlaid; (bottom left)
absolute magnitude against date by filter; (top right) the combined,
colour-corrected absolute magnitude light curve; (middle right) the
Lomb--Scargle periodogram, with the adopted peak marked and the masked
diurnal-alias regions shaded grey; and (bottom right) the light curve folded at
the adopted rotation period, with the running-average curve
and measured amplitude overlaid. For each
target we also present the convex shape and spin-state model derived by
lightcurve inversion; we treat the
Lomb--Scargle and convex-model rotation periods as independent determinations,
noting where they agree and, where they do not, carrying both as alternatives.

The derived parameters are organised into two tables. Table~\ref{tab:phot}
summarises the photometric quantities --- absolute magnitudes, phase slopes,
the \como{} colour, the inferred albedo class and our suggested taxonomy ---
together with published taxonomic classifications.
Table~\ref{tab:rotation} collects the rotational and shape properties --- the
Lomb--Scargle and convex-model periods, the light-curve amplitude, the implied
axis-ratio lower limit, the convex spin-state solution and published rotation
periods.

We use two derived quantities to inform the shape and taxonomic discussion. First, a
lower limit on the elongation follows from the light-curve amplitude
$\Delta m$,
\begin{equation}
  a/b \;\geq\; 10^{\,0.4\,\Delta m},
  \label{eq:axisratio}
\end{equation}
which is a strict lower bound because it neglects both the unknown aspect angle
and the amplitude--phase-angle relation \citep{zappala1990}; our
running-average amplitudes are themselves conservative
(Section~\ref{sec:methods_amplitude}), so the true elongations are larger.

Second, we use the shared phase slope
$G_{\mathrm{avg}} = \tfrac{1}{2}(G_o + G_c)$ as a coarse albedo indicator.
\citet{shevchenko2019} related the phase integral $q$ to the $H,G$ slope
through $q = 0.290 + 0.684\,G$ and tabulated $q$ for asteroids grouped by
geometric albedo. Inverting their relation on those groups gives characteristic
$H,G$ slopes of $G \approx 0.07$--$0.09$ for low-albedo ($p \approx 0.06$,
C-complex), $G \approx 0.19$--$0.23$ for moderate-albedo ($p \approx 0.20$,
S-complex), and $G \approx 0.37$--$0.50$ for high-albedo ($p \approx 0.45$,
E-complex) surfaces. We use these ranges to assign each target to a low-, moderate- or high-albedo
class from its $G_{\mathrm{avg}}$, and combine that albedo class with the
\como{} colour to infer a probable taxonomy. A neutral colour
($c-o \lesssim 0.26$) with a low albedo indicates a C-type; an intermediate
colour ($c-o \approx 0.26$--$0.35$) with a moderate-to-high albedo points to the
X complex --- an M-type at moderate albedo, or an E-type at high albedo; a red
colour ($c-o \approx 0.35$--$0.45$) with a moderate-to-high albedo indicates an
S-type; and an extremely red colour ($c-o \gtrsim 0.45$) at moderate albedo
indicates an L- or D-type. We stress that neither $G$ nor $(c-o)$ is a
substitute for spectroscopy: the two-parameter classification is suggestive
only, and is offered precisely so that spectroscopic follow-up (e.g. \cite{lee2025}) --- including the
\ema{} encounters themselves --- can test it.

\subsection{(10253) Westerwald}
\label{sec:res_westerwald}

\westerwald{} is the first flyby target (Figure~\ref{fig:westerwald}). We determine a colour of
$(c-o) = 0.41 \pm 0.01$~mag which is the reddest colour among the six flyby
targets and is typical of an S-type surface in the \atlas{} two-colour system
\citep{erasmus2020}. The phase slopes are correspondingly shallow,
$G_o = 0.37 \pm 0.02$ and $G_c = 0.31 \pm 0.03$ ($G_{\mathrm{avg}} = 0.34$),
consistent with a moderate-to-high albedo as expected for an S-type body and
reinforcing the colour-based classification. \cite{lee2025} also report an S-type classification via a visible-wavelength spectrum.

The periodogram signal is weak, with a low folded amplitude of $0.12$~mag
implying only a modest elongation ($a/b \gtrsim 1.11$; Eq.~\ref{eq:axisratio}).
This is consistent with a fairly spheroidal shape and/or a near-pole aspect
during much of the \atlas{} baseline; either way the small photometric
modulation makes the period harder to extract. We nonetheless adopt
$P_{\mathrm{LC}} = 1.81821 \pm 0.00002$~h, i.e.\
$P_{\mathrm{rot}} = 3.6364 \pm 0.0001$~h. Our period agrees closely with the
value of $3.6370 \pm 0.0001$~h independently reported by
\citet{pravec2024web}, which confirms this determination despite the weak
periodogram signal.

The convex shape model cannot be derived uniquely by lightcurve inversion
(Section~\ref{sec:methods_shapes}). The ATLAS photometric data set contains many points with large photometric errors, and searching for the best-fit period yields no unique solution. After removing points with errors larger than 0.2~mag and repeating the period search, we found the sideral rotation period $P = 3.636862 \pm 0.000004$~h which is in agreement with our periodogram derived period. However, there are many models with this rotation period but with different spin-axis orientations that fit the data to the same level of residuals. The only constraint we can put on the spin-axis orientation is that the rotation is retrograde.


\subsection{(623) Chimaera}
\label{sec:res_chimaera}

\chimaera{} is the largest remnant of the primitive C-type Chimaera family and
the second-largest \ema{} target (Figure~\ref{fig:chimaera}). We measure a colour of
$(c-o) = 0.24 \pm 0.01$~mag --- a colour very typical of a C-type asteroid in
the \atlas{} system \citep{erasmus2020}. This is corroborated by the steep
phase slope, $G_{\mathrm{avg}} = 0.13$, among the
lowest in our sample and indicative of a dark, low-albedo surface, and consistent with the
Xc or C classification obtained from visible spectroscopy by
\citet{morate2019} and \citet{lee2025}, together with the low albedo reported in the LCDB of $\sim$0.04 from several independent sources.

The periodogram does not show a single clean period; the three strongest
light-curve peaks are at $10.524$, $7.311$ and $5.601$~h. We report the
strongest, giving $P_{\mathrm{rot}} = 21.047 \pm 0.001$~h and the folded amplitude
is low ($0.13$~mag; $a/b \gtrsim 1.13$). We note that the LCDB lists a rotation
period of $14.635$~h \citep[U${=}3$;][]{fleenor2007,warner2009}, which does not
correspond to our strongest peak but is very close to twice our second-strongest
peak ($2\times7.311 = 14.62$~h). Visual inspection of the light curve published by \cite{fleenor2007} shows a convincing dense light-curve so we therefore believe the second strongest periodogram peak associated with a $P_{\mathrm{rot}} = 14.62$~h to be the most likely period.  

\begin{figure}
\centering
\includegraphics[width=\columnwidth, trim=2cm 6.5cm 1cm 0.8cm, clip]{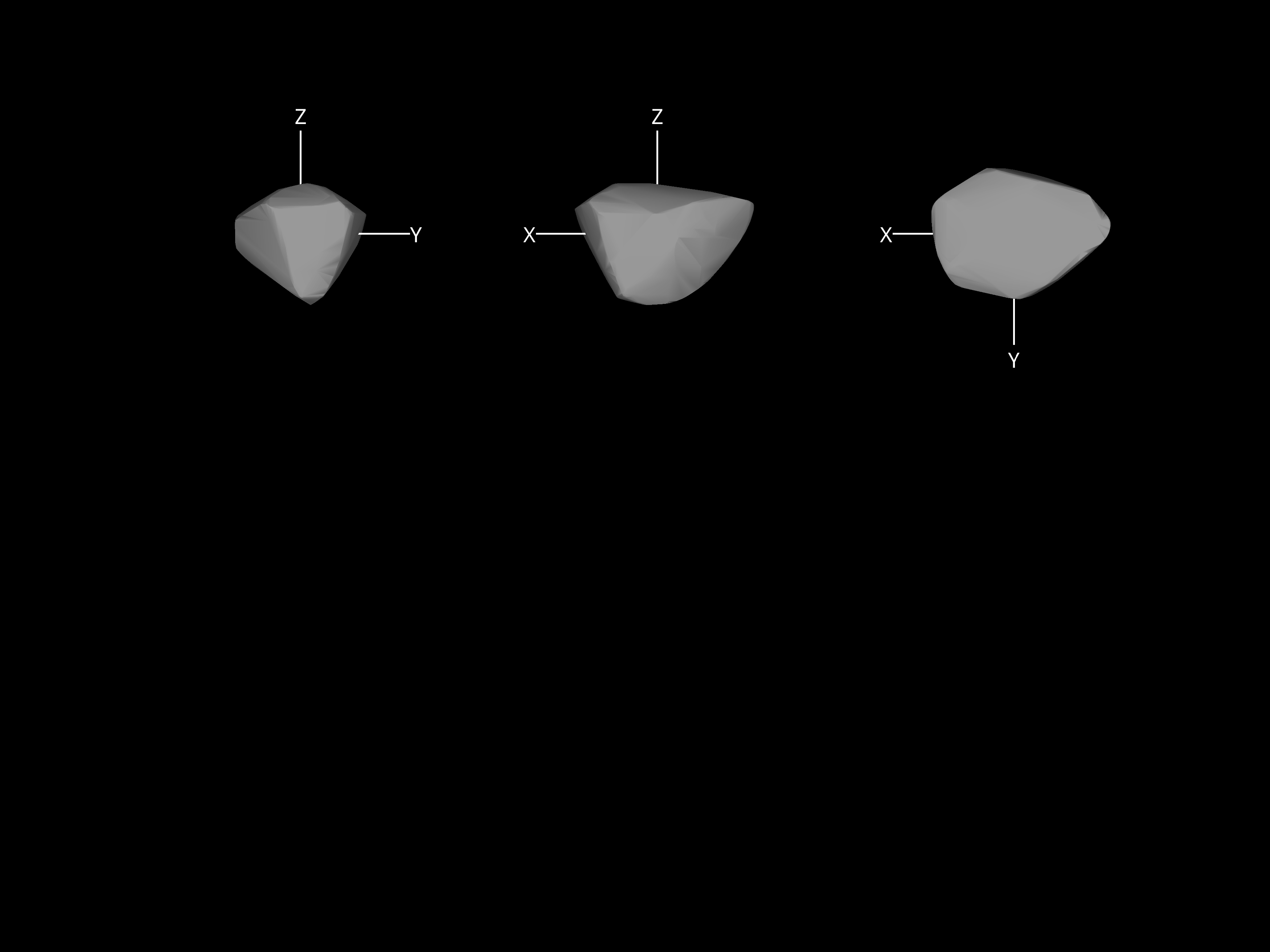}\\
\includegraphics[width=\columnwidth, trim=2cm 6.5cm 1cm 0.8cm, clip]{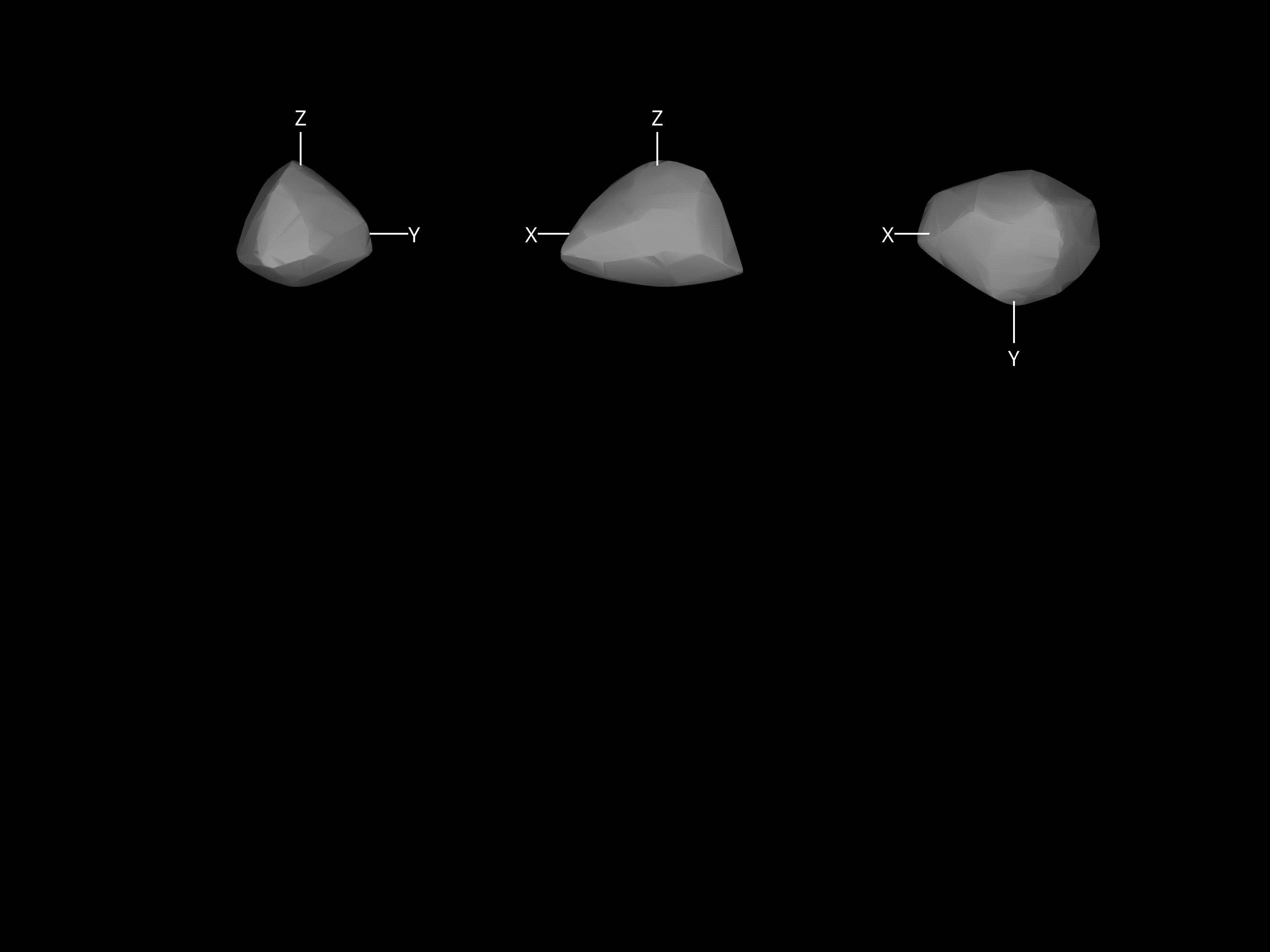}
\caption{Convex shape
models of \chimaera{} from \atlas{} photometry, shown in three orthogonal
projections. The top model (our preferred model) corresponds to the pole direction $(128^{\circ}, 2^{\circ})$, the bottom one to $(309^{\circ}, 2^{\circ})$.}
\label{fig:shape_chimaera}
\end{figure}

The convex shape models derived by lightcurve inversion
(Section~\ref{sec:methods_shapes}) are shown in Figure~\ref{fig:shape_chimaera}. There are two models that provide the same level of fit to photometric data. They have the same sidereal rotation period of $P = 14.62541 \pm 0.00003$~h (which agrees with \citet{fleenor2007}) and spin axis directions of $(128 \pm 1^{\circ}, 2 \pm 2^{\circ})$ and $(309 \pm 1^{\circ}, 2 \pm 2^{\circ})$, respectively. The color $c - o$ is $0.289 \pm 0.005$~mag and $0.286 \pm 0.005$~mag for the first and the second pole solution, respectively, is slightly redder than the derived colour using the method described in Section~\ref{sec:methods_colour} and perhaps suggests a Xc over a C-type. The first model agrees with the independent solution of \cite{Dur.Han:23} that was obtained from Gaia DR3 data.

To resolve the ambiguity in the pole direction, we used stellar occultation data collected by D.~Herald in the Occult software.\footnote{\url{https://occultations.org/sw/occult/V4_2026_01_01html.html}} For \chimaera{}, there were two stellar occultations on 22 February 2019 (3 positive observations) and 24 May 2024 (4 positive observations) that we compared with the projected silhouettes of our models using the approach of \cite{Dur.ea:11}. The model with the pole direction $(128^{\circ}, 2^{\circ})$ provided a significantly better fit than the other model, so we preferred this solution. The volume-equivalent diameter of the model that provided the best fit was 36~km.

\subsection{(13294) Rockox}
\label{sec:res_rockox}

\rockox{} yields the cleanest rotational detection in the sample
(Figure~\ref{fig:rockox}) and the colour of $(c-o) = 0.30 \pm 0.02$~mag is intermediate between the C- and S-complex loci and is therefore consistent with an
X-type (moderate) visible spectral slope. The moderate phase slope,
$G_o = 0.24 \pm 0.02$ ($G_{\mathrm{avg}} = 0.20$), does not by itself break the
degeneracy, since the X complex spans a wide albedo range but does suggest possible M-type classification. \citet{lee2025} also classifies \rockox{} as X and its measured NEOWISE albedo ($p_V = 0.14 \pm 0.02$; \citealt{mainzer2019}) is consistent with the moderate-albedo X/M-type surface we
infer.

The rotation period is unambiguous: a strong, isolated peak at
$P_{\mathrm{LC}} = 1.62130 \pm 0.00002$~h gives
$P_{\mathrm{rot}} = 3.24261 \pm 0.00003$~h. This agrees closely with published
determinations of $3.24305$~h \citep{durech2020,pal2020} and $3.24298$~h \citep{pravec2024web}. The $3.137$~h value of \citet{erasmus2020} is very likely a one-day alias of the true period. The folded light curve is clearly bimodal with the largest amplitude in the sample,
$0.46$~mag, implying a substantially elongated body with
$a/b \gtrsim 1.53$ (Eq.~\ref{eq:axisratio}).

\begin{figure}
\centering
\includegraphics[width=\columnwidth, trim=2cm 6.5cm 1cm 0.8cm, clip]{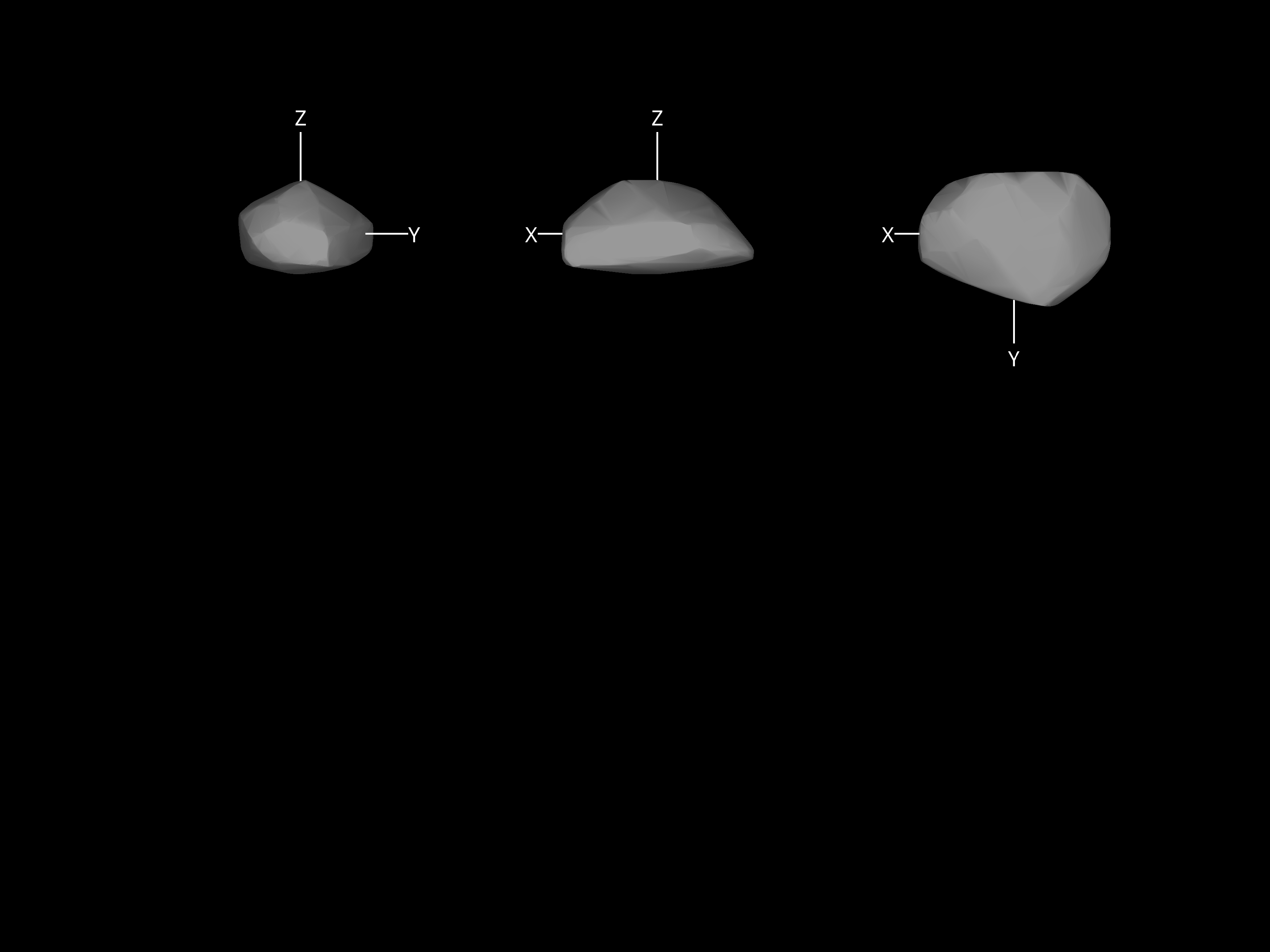}
\caption{Convex shape model of \rockox{} from \atlas{} photometry, shown in three orthogonal projections.}
\label{fig:shape_rockox}
\end{figure}

The convex shape model derived by lightcurve inversion
(Section~\ref{sec:methods_shapes}) is shown in Figure~\ref{fig:shape_rockox}.
The inversion converged on a sidereal period of $3.2429635 \pm 0.0000005$~h, in excellent
agreement with our Lomb--Scargle and the
published values; this mutual consistency
between two independent methods makes \rockox{} the most securely determined
period in the sample. The best-fit spin pole lies at ecliptic coordinates
$(\lambda, \beta) = (133 \pm 2^{\circ}, -76 \pm 2^{\circ})$, indicating a retrograde
rotation. The convex
model has $a/b \approx 1.45$ (computed from a dynamically equivalent ellipsoid), which is similar  to the
$a/b \gtrsim 1.53$ lower limit implied by the light-curve amplitude. The color index from light curve inversion is $c - o = 0.315 \pm 0.005$~mag and agrees with an X-complex classification.

\subsection{(88055) Ghaf}
\label{sec:res_ghaf}

For \ghaf{} (Figure~\ref{fig:ghaf}) we determine a colour of  $(c-o) = 0.34 \pm 0.02$~mag. This is slightly
bluer than the canonical S-type colour but is still most consistent with an
S-type classification. The shallow phase slopes, $G_o = 0.36 \pm 0.04$ and
$G_c = 0.28 \pm 0.06$ ($G_{\mathrm{avg}} = 0.32$), support a moderate-to-high
albedo and hence the S-type assignment, though the uncertainties are the
largest in the sample as this is the faintest target. \ghaf{} was included in
the MOVIS near-infrared colour survey of \citet{popescu2018} but its colours did
not yield a definitive type (hence their U flag in the LCDB). That said, their two algorithms tentatively favour the K/L group and their MOVIS near-infrared colours (($Y-J = 0.28$, $J-Ks = 0.38$)) fall within the S and K-type locus, so are broadly compatible with, and mildly supportive of, our S-type classification. On the contrary, \cite{lee2025} measure a distinctly blue spectral slope and assign a B-type taxonomy and \cite{mainzer2019} report an albedo $p_V = 0.054 \pm 0.006$ which supports low-albedo carbonaceous asteroid. That independent measurements both support and contradict our S-type assignment is difficult to reconcile and may hint at genuine surface heterogeneity and \ghaf{} is therefore a clear priority for dedicated follow-up ahead of encounter.  

We adopt $P_{\mathrm{LC}} = 4.78385 \pm 0.00016$~h, i.e.\
$P_{\mathrm{rot}} = 9.5677 \pm 0.0003$~h. The periodogram peak is not
especially strong, but the folded and phased data are reasonably coherent, with
an amplitude of $0.21$~mag ($a/b \gtrsim 1.21$). \citet{pravec2024web} report a
period of $5.918 \pm 0.001$~h for \ghaf{} (their quality code U${=}2$), which
differs from our $9.5677$~h and may be related to it by diurnal aliasing. As
both determinations are uncertain --- the \citet{pravec2024web} value is flagged
as tentative and our periodogram peak is not strong --- we regard the rotation
period of \ghaf{} as unresolved. We could also not derive a reliable shape model for \ghaf{}.


\subsection{(23871) Ousha}
\label{sec:res_ousha}

\ousha{} is the most poorly constrained target, with the fewest usable epochs
(Figure~\ref{fig:ousha}). We derive $(c-o) = 0.29 \pm 0.03$~mag --- again an
intermediate colour that is not conclusively C or S, and most consistent with
an X-like visible slope but with uncertainty in the colour large enough that we can not rule out a C-like object. The per-filter slopes are strongly discrepant,
$G_o = 0.56 \pm 0.08$ against $G_c = 0.11 \pm 0.05$; we attribute this to the
poor data quality but perhaps this could be due to some unique intrinsic property. The low $G_c$ suggests a low-albedo object while the high $G_o$ suggests a high-albedo object. The 
high $G_{\mathrm{avg}}$ that we employ for classification with the neutral $(c-o)$ colour that we determine raises the possibility of an E-type (high-albedo X-complex) classification for \ousha{}, which we flag albeit tentatively. \cite{lee2025} show a clear B-type spectrum which is contradictory to our tentative high-albedo E-type classification (unless one favours our fitted $G_c$ over the fitted $G_o$) and therefore \ousha{} should also be  priority for dedicated follow-up ahead of encounter to resolve this discrepancy.

The rotation period is not clear-cut either. Our strongest peak gives a nominal
$P_{\mathrm{rot}} = 20.52 \pm 0.003$~h ($P_{\mathrm{LC}} = 10.262$~h), but with
low confidence. \citet{pravec2024web} report $8.3515$~h (their quality code
U${=}3$), and our third-strongest peak ($P_{\mathrm{LC}} = 13.678$~h) is the
one-day alias of that period; this leads us to prefer the \citet{pravec2024web} value as the more probable rotation period. The measured LC
amplitude at our nominal period is $0.39$~mag ($a/b \gtrsim 1.43$), but this is unreliable given the period ambiguity.

The ATLAS data alone were not sufficient to provide a shape model and therefore
a unique solution for the rotation period. However, sprinkling in data from Pan-STARRS,
Catalina, Mt. Lemmon, and Gaia DR3 made the inversion converged on a sidereal
period of $\sim$$8.35$~h which matches the period reported by
\citet{pravec2024web}.


\subsection{(59980) Moza}
\label{sec:res_moza}

For \moza{} (Figure~\ref{fig:moza}) we measure an almost perfect S-type colour of $(c-o) = 0.37 \pm 0.01$~mag in the \atlas{}-filter system \citep{erasmus2020}. The
moderate phase slopes, $G_o = 0.26 \pm 0.02$ and $G_c = 0.18 \pm 0.03$
($G_{\mathrm{avg}} = 0.22$), are consistent with the intermediate albedo of an
S-type body. The LCDB lists a K classification for \moza{} \citep{warner2009}; this is an
assumed class based on membership of the (predominantly K-type) Eos family
rather than a measured taxonomy, and so does not independently corroborate our assignment. The measured NEOWISE albedo ($p_V = 0.12 \pm 0.02$; \citealt{mainzer2019}) is nonetheless consistent with the moderate-albedo S classification we infer. \cite{lee2025} also report a S-type spectrum (but flag a possible L-type too).

The rotation period is convincing: $P_{\mathrm{LC}} = 2.18143 \pm 0.00003$~h
gives $P_{\mathrm{rot}} = 4.36286 \pm 0.00006$~h, with a coherent folded
modulation of $0.27$~mag ($a/b \gtrsim 1.28$). Our period is independently
confirmed by \citet{pravec2024web}, who report $4.3647 \pm 0.0005$~h (quality
code U${=}3$). The earlier value of $136 \pm 40$~min ($2.27 \pm 0.67$~h) from
\citet{erasmus2018} is, after visual inspection of their published dense light
curve --- which spans only $\sim$1~h and does not capture two full minima ---
best regarded as a lower limit on the period rather than a conflicting
determination, and is compatible with our $\sim$4.36~h rotation period.

\begin{figure}
\centering
\includegraphics[width=\columnwidth, trim=2cm 6.5cm 1cm 0.8cm, clip]{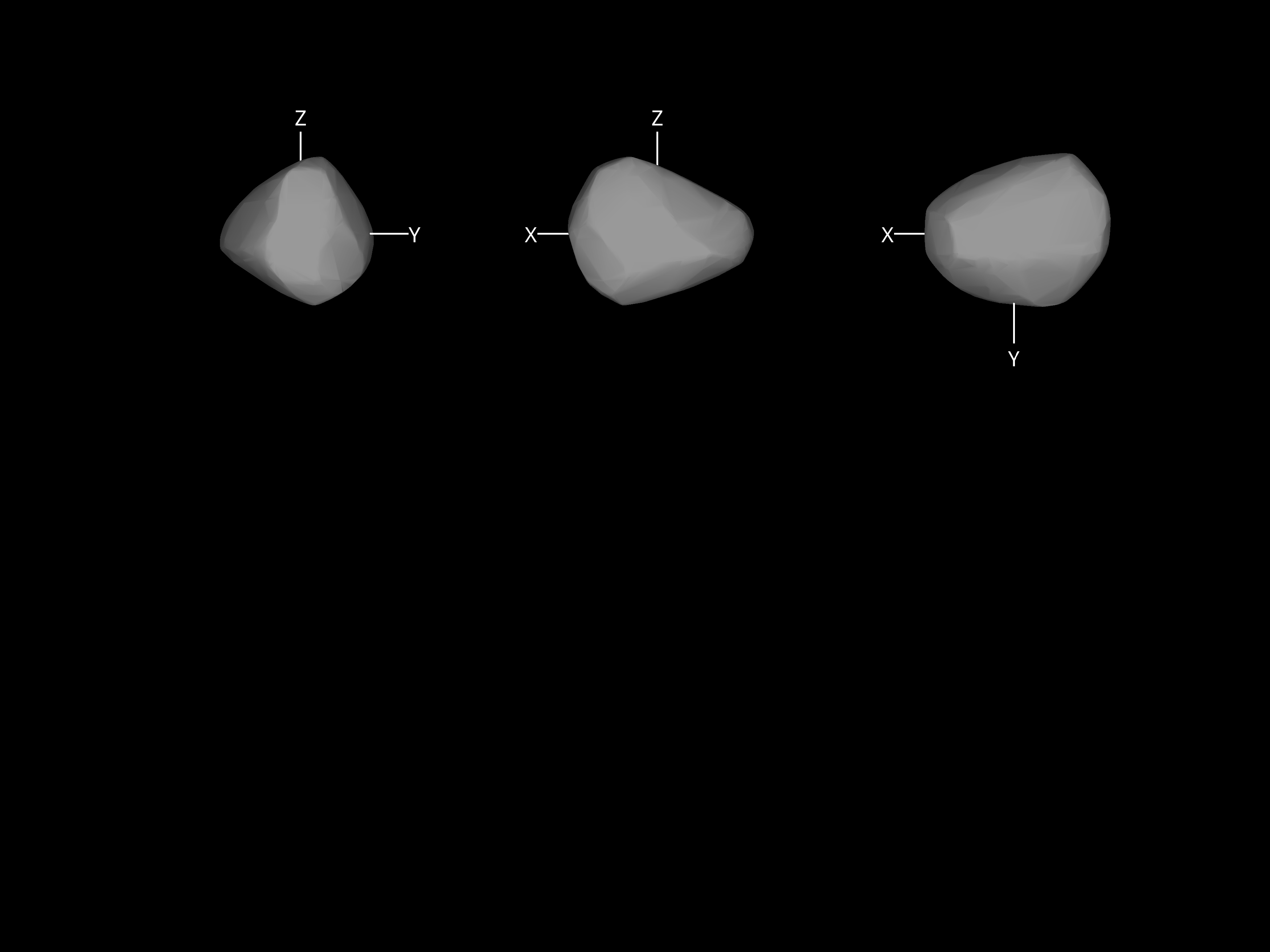}\\
\includegraphics[width=\columnwidth, trim=2cm 6.5cm 1cm 0.8cm, clip]{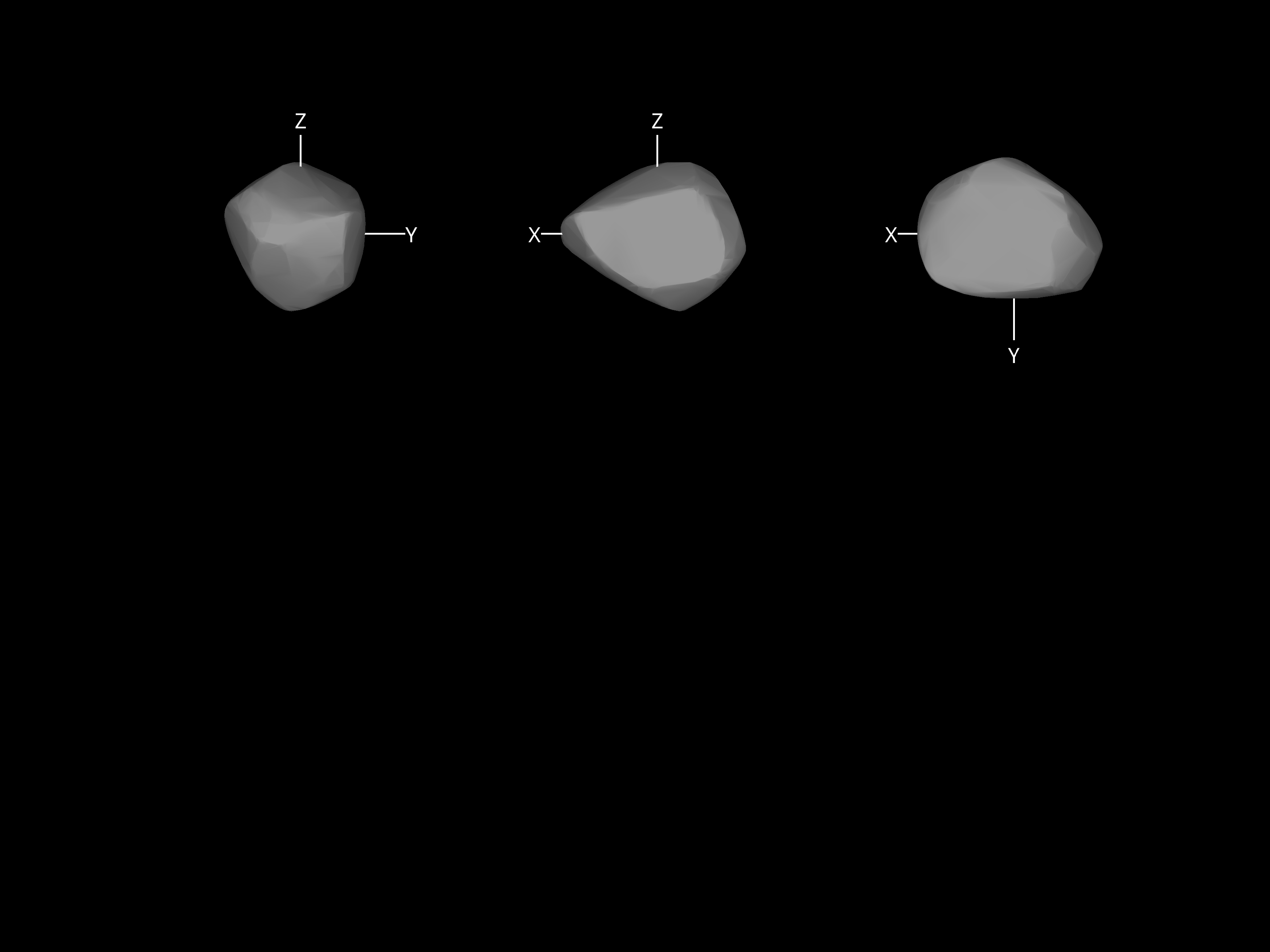}
\caption{Convex shape
models of \moza{} from \atlas{} photometry, shown in three orthogonal
projections. The top model corresponds to the pole direction $(25^{\circ}, 49^{\circ})$, the bottom one to $(190^{\circ}, 67^{\circ})$.}
\label{fig:shape_moza}
\end{figure}

The convex shape models derived by lightcurve inversion
(Section~\ref{sec:methods_shapes}) are shown in Figure~\ref{fig:shape_moza}. There are two solutions for the pole: $(25 \pm 3^{\circ}, 49 \pm 4^{\circ})$ or $(190 \pm 3^{\circ}, 67 \pm 3^{\circ})$ with almost identical sidereal rotation and colour values measured compared to our other method.

\subsection{(269) Justitia}
\label{sec:res_justitia}

\justitia{} is the rendezvous and landing target of \ema{} and, being both the
largest and brightest, the best-observed object in our sample
(Figure~\ref{fig:justitia}). We measure the
reddest colour in our sample ($c-o = 0.48 \pm 0.01$~mag) which is redder than the typical Trojan colour of
$\sim$0.4 measured in the \atlas{}-filter system by \citet{mcneill2021}, although that
study did include some Trojans with \atlas{} colours as red as
$\sim$0.5--0.55. The extreme colour is fully consistent with the well
established, exceptionally steep visible--near-infrared spectral slope of
\justitia{}, which exceeds that of any D-type asteroid and resembles the very
red Centaurs and trans-Neptunian objects, and which has been interpreted as
evidence for an origin in the outer Solar System followed by inward migration
\citep{hasegawa2021,marciniak2025}. \justitia{} is spectroscopically classified
as a D-type in the Bus--DeMeo system \citep[or Ld in the SMASS
system;][]{humes2024,busbinzel2002,lee2025}, with a visible spectral slope steeper than
any other D-type main-belt asteroid \citep{humes2024}, consistent with the very
red, featureless slope we recover. The moderate phase slopes,
$G_o = 0.21 \pm 0.01$ and $G_c = 0.25 \pm 0.01$
($G_{\mathrm{avg}} = 0.23$), suggests a higher albedo than the low geometric albedo
($\sim$0.06--0.08) reported from occultation and thermophysical modelling
\citep{marciniak2025}.

Our rotation period, $P_{\mathrm{rot}} = 33.0997 \pm 0.0029$~h (from
$P_{\mathrm{LC}} = 16.54984 \pm 0.00145$~h), is in agreement with the
published value of $33.12962$~h derived by \citet{marciniak2025} from dense
light curves. The folded amplitude is $0.185$~mag, implying a
lower limit on the elongation of $a/b \gtrsim 1.19$; this is a strict lower bound.

\begin{figure}
\centering
\includegraphics[width=\columnwidth, trim=2cm 6.5cm 1cm 0.8cm, clip]{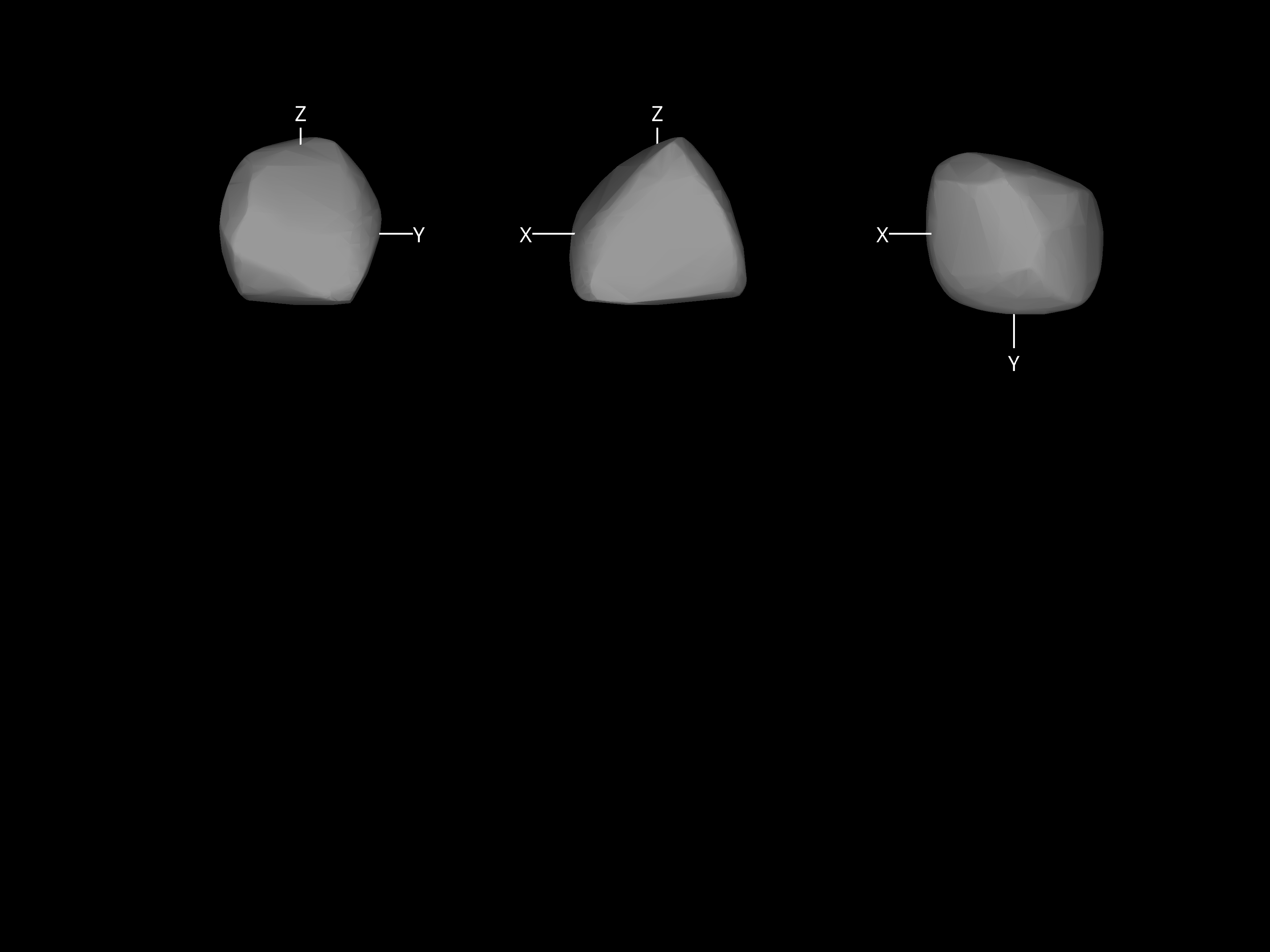}\\
\includegraphics[width=\columnwidth, trim=2cm 6.5cm 1cm 0.8cm, clip]{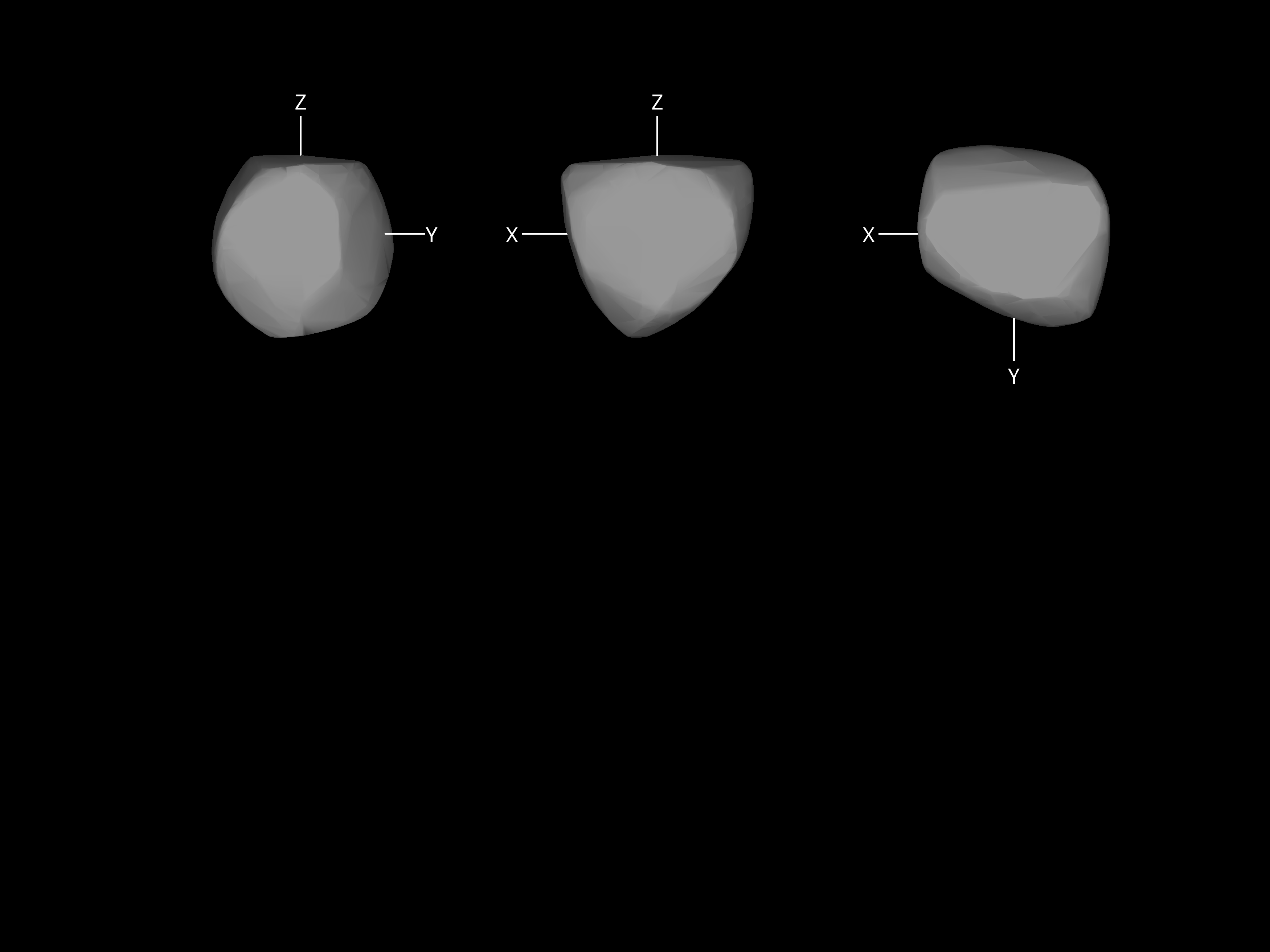}
\caption{Convex shape
models of \justitia{} from \atlas{} photometry, shown in three orthogonal
projections. The top model corresponds to the pole direction $(254^{\circ}, -53^{\circ})$, the bottom one (our preferred model) to $(73^{\circ}, -63^{\circ})$.}
\label{fig:shape_justitia}
\end{figure}

Our own convex shape models derived by lightcurve inversion are shown in Figure~\ref{fig:shape_justitia}. The two possible pole directions $(254 \pm 2^{\circ}, -53 \pm 1^{\circ})$ and $(73 \pm 2^{\circ}, -63 \pm 1^{\circ})$ are similar to values derived by \cite{marciniak2025}; the sidereal rotation period $P = 33.12970 \pm 0.00004$~h matches our period derived via LS. The shape models published by \citep{marciniak2025} were reconstructed from a joint inversion of optical light curves and thermal infrared data, which makes them more regular than our models derived from sparse ATLAS photometry.

The pole ambiguity can be resolved by comparing the projected silhouettes of the two models with the stellar occultation observed on 31 August 2023 and analyzed by \cite{buie2025}. Our model with the pole direction $(73^{\circ}, -63^{\circ})$ is preferred as it provides a better agreement with the observed chords than the other model. The best-fit volume-equivalent diameter is 53~km. The best match is given by the model with pole direction $(73^{\circ}, -81^{\circ})$ by \cite{marciniak2025}. They determined the size from thermal infrared data to $58 \pm 2$~km, and scaling their model to fit the occultation yields the same size. 


\subsection{Summary of derived parameters}
\label{sec:res_table}

\begin{figure}
\centering
\includegraphics[width=\columnwidth]{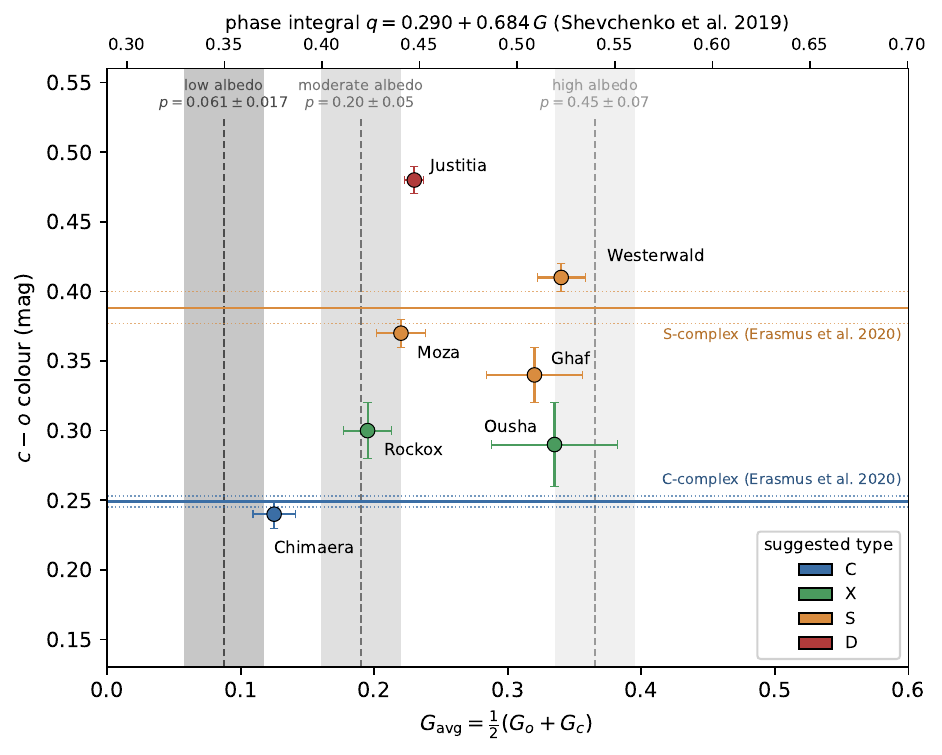}
\caption{Taxonomic summary of the seven EMA targets in the
    $(c-o)$ colour versus mean phase-slope
    $G_{\mathrm{avg}} = \tfrac{1}{2}(G_o + G_c)$ plane from the values in Table \ref{tab:phot}. Points are coloured by
    suggested taxonomic type (C, X, S, D; legend). Horizontal solid lines
    mark the typical median S- and C-complex $c-o$ colours from
    \citet{erasmus2020}, with dotted lines showing their quoted uncertainties.
    Vertical dashed lines and grey bands mark the low-, moderate-, and
    high-albedo classes of \citet{shevchenko2019}, placed at their tabulated
    phase integrals $q$ derived from $G$ via $q = 0.290 + 0.684\,G$; the
    corresponding geometric albedos $p$ are annotated above. The top axis gives
    the phase integral $q$ on the same scale.}
\label{fig:taxonomy}
\end{figure}

The full set of derived photometric and rotational parameters is collected in
Tables~\ref{tab:phot} and~\ref{tab:rotation}, and the colour--albedo
distribution with our suggested taxonomies is summarised in
Figure~\ref{fig:taxonomy}. Taken together, the seven targets span most of the
main taxonomic groupings. On the two-parameter $(c-o)$ versus $G_{\mathrm{avg}}$
plane, \chimaera{} falls cleanly in the low-albedo, neutral-colour C-type
region; \westerwald{}, \ghaf{} and \moza{} occupy the moderate-to-high-albedo,
red-colour S-type region; \rockox{} and \ousha{} lie in the intermediate
X-complex zone; and \justitia{} sits apart from all of them, its extreme red
colour ($c-o = 0.48$) placing it firmly in the L/D domain. The albedo classes
inferred from $G_{\mathrm{avg}}$ via the \citet{shevchenko2019} relation agree
with the measured NEOWISE albedos for four out of five targets for which these exist
(\chimaera{}, \rockox{}, \moza{} and \justitia{}; Section~\ref{sec:results}),
supporting the reliability of the phase-slope-based albedo estimate. Curiously, the only mismatches in taxonomy between
this work and those published is for the two objects \cite{lee2025} classify as B-types, which we classify as an S-type  for \ghaf{} and X-type for \ousha{} (albeit a tentative classification for \ousha{} where we can't rule out a low-albedo C-like classification). There is no obvious reason why our $(c-o)$ colours would not
correctly identify very blue B-type spectra, but \cite{lee2025} visible spectra clearly show negative slopes in reflectance consistent with B-type asteroids. Our derived relatively high albedo for these two objects is also in contradiction with a B-type classification.

The rotation periods (Table~\ref{tab:rotation}) span an order of magnitude,
from the $\sim$3.2~h fast rotator \rockox{} to the $\sim$33~h slow rotator
\justitia{}. Five periods --- those of \westerwald{}, \chimaera{} \rockox{}, \moza{} and
\justitia{} --- are secure, being either well isolated in the periodogram or converge in the convex model and are in
agreement with independent published determinations. One, \ousha{}{}, is not fully resolved from our analysis but our data does support the published period by \cite{pravec2024web}.
The period of \ghaf{} is unresolved, our value and that of
\citet{pravec2024web} disagreeing at a level consistent with diurnal aliasing.

\begin{table*}
\centering
\caption{Photometric parameters and taxonomic inference, in \mbr{}
encounter order. $H_o$ and $H_c$ are the per-filter absolute magnitudes from
the independent $H,G$ fits; $(c-o)$ is computed on the shared
$G_{\mathrm{avg}}=\tfrac{1}{2}(G_o+G_c)$ slope and not from $H_c-H_o$ 
(see Section~\ref{sec:methods_colour}). The albedo class is inferred from
$G_{\mathrm{avg}}$ via the \citet{shevchenko2019} phase-integral relation, and
the suggested taxonomy from the albedo class and $(c-o)$ together
(Section~\ref{sec:results}).}
\label{tab:phot}
\setlength{\tabcolsep}{4pt}
\begin{tabular}{lcccccccc}
\toprule
Target & $H_o$ & $H_c$ & $(c-o)$ & $G_o$ & $G_c$ & Albedo & Tax. & Tax.\\
       & (mag) & (mag) & (mag) & & & class & (ours) & (pub.)\\
\midrule
\westerwald{} & $15.16\pm0.02$ & $15.53\pm0.02$ & $0.41\pm0.01$ & $0.37\pm0.02$ & $0.31\pm0.03$ & mod.--high & S & S$^{\dagger}$ \\
\chimaera{} & $10.60\pm0.02$ & $11.00\pm0.03$ & $0.24\pm0.01$ & $0.07\pm0.01$ & $0.18\pm0.03$ & low & C & Xc$^{\ddagger}$, X or C$^{\dagger}$ \\
\rockox{} & $14.22\pm0.03$ & $14.42\pm0.03$ & $0.30\pm0.02$ & $0.24\pm0.02$ & $0.15\pm0.03$ & mod. & X (M) & X$^{\dagger}$ \\
\ghaf{} & $15.20\pm0.03$ & $15.47\pm0.05$ & $0.34\pm0.02$ & $0.36\pm0.04$ & $0.28\pm0.06$ & mod.--high & S & B$^{\dagger}$ \\
\ousha{} & $15.29\pm0.06$ & $15.17\pm0.06$ & $0.29\pm0.03$ & $0.56\pm0.08$ & $0.11\pm0.05$ & mod.--high & X (E)$^{\mathrm{a}}$ & B$^{\dagger}$ \\
\moza{} & $13.15\pm0.02$ & $13.45\pm0.03$ & $0.37\pm0.01$ & $0.26\pm0.02$ & $0.18\pm0.03$ & mod. & S & K$^{\ast,\S}$, S or L$^{\dagger}$ \\
\justitia{} & $9.48\pm0.01$ & $10.00\pm0.01$ & $0.48\pm0.01$ & $0.21\pm0.01$ & $0.25\pm0.01$ & mod. & D & D$^{\P}$, D$^{\dagger}$ \\
\bottomrule
\end{tabular}
\vspace{2pt}
\begin{minipage}{\textwidth}
\footnotesize
\textbf{References.}
$^{\dagger}$~\citet{lee2025};
$^{\ddagger}$~\citet{morate2019};
$^{\S}$~\citet{warner2009};
$^{\P}$~\citet{humes2024}.\\
$^{\ast}$~Assumed LCDB class, based on family or orbital-group membership rather
than a spectroscopic or colour measurement, and so not an independent
classification.\\
$^{\mathrm{a}}$~This is a tentative classification and we cannot rule out a low albedo C/B-like object that is reported by \cite{lee2025} (see Section \ref{sec:res_ousha}).\\
\end{minipage}
\end{table*}

\begin{table*}
\centering
\caption{Rotational and shape parameters, in \mbr{} encounter order.
$P_{\mathrm{LS}}$ is the rotation period from the Lomb--Scargle analysis
(twice the light-curve period; Section~\ref{sec:methods_period});
$P_{\mathrm{conv}}$ is the sidereal period from convex inversion
(Section~\ref{sec:methods_shapes}). The amplitude is the peak-to-peak of the
running-average folded light curve (a lower limit), and $a/b$ the implied
elongation lower limit (Eq.~\ref{eq:axisratio}). $(\lambda,\beta)$ is the
convex-model pole. Published rotation periods are listed with their source.}
\label{tab:rotation}
\setlength{\tabcolsep}{4pt}
\begin{tabular}{lcccccl}
\toprule
Target & $P_{\mathrm{LS}}$ & $P_{\mathrm{conv}}$ & Amp. & $a/b$ & $(\lambda,\beta)$ & $P_{\mathrm{rot}}$ (pub.) \\
       & (h) & (h) & (mag) & $\geq$ & (\degr) & (h) \\
\midrule
\westerwald{} & $3.6364\pm0.0001$ & $3.636862 \pm 0.000004$ & $0.12$ & $1.11$ & retrograde rotation$^{\mathrm{c}}$ & $3.6370$$^{\dagger}$ \\
\chimaera{} & $21.047\pm0.001^{\mathrm{a}}$ & $14.62541 \pm 0.00003$  & $0.13$ & $1.13$ & $(128 \pm 1^{\circ}, 2 \pm 2^{\circ})^{\mathrm{d}}$ or $(309 \pm 1^{\circ}, 2 \pm 2^{\circ})$ & $14.635$$^{\ddagger}$ \\
\rockox{} & $3.24261\pm0.00003$ & $3.2429635 \pm 0.0000005$ & $0.46$ & $1.53$ & $(133 \pm 2^{\circ}, -76 \pm 2^{\circ})$ & $3.243$$^{\S,\P,\dagger}$ \\
\ghaf{} & $9.5677\pm0.0003$ &  & $0.21$ & $1.21$ &  & $5.918$$^{\dagger}$ \\
\ousha{} & $20.523\pm0.003^{\mathrm{b}}$ &  & $0.39$ & $1.43$ &  & $8.3515$$^{\dagger}$ \\
\moza{} & $4.36286\pm0.00006$ & $4.362456 \pm 0.000002$ & $0.27$ & $1.28$ & $(25 \pm 3^{\circ}, 49 \pm 4^{\circ})$ or $(190 \pm 3^{\circ}, 67 \pm 3^{\circ})$ & $4.3647$$^{\dagger}$ \\
\justitia{} & $33.0997\pm0.0029$ & $33.12970 \pm 0.00004$ & $0.19$ & $1.19$ & $(254 \pm 2^{\circ}, -53 \pm 1^{\circ})$ or $(73 \pm 2^{\circ}, -63 \pm 1^{\circ})^{\mathrm{d}}$ & $33.12962$$^{\|}$ \\
\bottomrule
\end{tabular}
\vspace{2pt}
\begin{minipage}{\textwidth}
\footnotesize
\textbf{References.}
$^{\dagger}$~\citet{pravec2024web};
$^{\ddagger}$~\citet{fleenor2007};
$^{\S}$~\citet{durech2020};
$^{\P}$~\citet{pal2020};
$^{\|}$~\citet{marciniak2025}.\\
$^{\mathrm{a}}$~We consider our second-strongest periodogram peak, corresponding
to $P_{\mathrm{rot}} = 14.62$~h, and our period derived from the convex shape model work ($14.62541$), to be the most likely period, in agreement with
\citet{fleenor2007}.\\
$^{\mathrm{b}}$~Our Lomb--Scargle period is likely a diurnal alias; our
third-strongest peak ($13.68$~h) is the one-day alias of the
\citet{pravec2024web} value, which we consider the more probable period (see also last paragraph in Section \ref{sec:res_ousha}).\\
$^{\mathrm{c}}$~We were unable to derive a unique model and the only constraint we can put on the spin-axis orientation is that
the rotation is retrograde.\\
$^{\mathrm{d}}$~Taking into account stellar occultation data, this is our preferred solution.\\
\end{minipage}
\end{table*}

%% file: sections/06_discussion.tex
\section{Discussion}
\label{sec:discussion}

\subsection{Taxonomy as pre-encounter predictions}
\label{sec:disc_taxonomy}

The \como{} colours and $G_{\mathrm{avg}}$-based albedo classes
(Section~\ref{sec:results}, Table~\ref{tab:phot}) spread the seven targets
across most of the main taxonomic groupings, from the dark, neutral C-type
\chimaera{} to the exceptionally red \justitia{}. The mission's 
expectation is that five of the seven are C-complex or otherwise primitive
\citep{alsaeed2025}; our colours support that unambiguously only for
\chimaera{}, leave the two X-complex candidates (\rockox{}, \ousha{})
compositionally ambiguous, and place our three S-type candidates
(\westerwald{}, \ghaf{}, \moza{}) outside the C-complex altogether (although other published works claim \ghaf{} to be a primitive type).

We do not read this as a contradiction. ATLAS two-band photometry is a coarse
discriminant since $(c-o)$ does not cleanly separate all classes and albedo inferred via phase slopes can be noisy
for the fainter targets. That said, collisional families are known to contain interlopers and our ground-based measurement that disagrees with a
family-based prior is exactly what motivates \emph{in situ} follow-up: where our
colours and the prior diverge, the EMA encounters will be decisive. We
therefore present these classifications as testable pre-encounter predictions
rather than final answers, and flag \ghaf{} and \ousha{} the least secure
classifications and those most in need of confirmation.

\subsection{The exceptional colour of (269) Justitia}
\label{sec:disc_justitia}

\justitia{} is the reddest object in our sample by a clear margin
($c-o = 0.48$ against $\leq 0.41$ for every other target), redder even than the
typical Jupiter Trojan measured in the same system \citep{mcneill2021}. Because
this colour is built from roughly a decade and hundreds of epochs of photometry
(Section~\ref{sec:methods_colour}), it establishes what spectroscopy alone
cannot: that the extreme spectral slope reported by \citet{hasegawa2021} and
\citet{marciniak2025} is an intrinsic, rotationally- and temporally-stable
property of the surface, reinforcing the interpretation of \justitia{} as a body
plausibly implanted from the outer Solar System. The moderate phase slope we
recover ($G_{\mathrm{avg}} = 0.23$) sits uneasily with the low geometric albedo
from occultation and thermophysical modelling \citep{marciniak2025,buie2025},
deepening rather than resolving the puzzle around this object.

\subsection{Rotation and shape}
\label{sec:disc_rotation}

The recovered periods span an order of magnitude, and five out of 7 are secure, each agreeing with independent
determinations. \rockox{} period agrees across four independent methods, making it the most
securely determined in the sample. The remaining two are the expected casualties
of residual diurnal aliasing: for \ousha{} we adopt the $\sim$8.35~h period of
\citet{pravec2024web}, of which our own third-strongest peak is the one-day
alias, and the period of \ghaf{} remains unresolved.

Convex inversion yields a unique shape and spin-state model for \rockox{}
and two pole solutions for \chimaera{}, \moza{} and
\justitia{}. For \chimaera{} and \justitia{} we indicate our preferred pole solution out of the two possibilities. No model was possible for \ghaf{}, \ousha{} or \westerwald{},
though \westerwald{}'s spin axis is constrained to be retrograde. The
dynamically-equivalent-ellipsoid axis ratios ($a/b = 1.45$ for \rockox{}, $1.38$
for \chimaera{}, $1.3$ for \moza{} and $1.1$ for \justitia{})
supersede the amplitude-based lower limits of Section~\ref{sec:results} and
confirm that \rockox{} is the most elongated target and \justitia{} among the
roundest.

\subsection{Implications for EMA}
\label{sec:disc_ema}

The parameters derived here bear directly on encounter planning: periods and
pole solutions constrain the sub-spacecraft geometry and illuminated hemisphere
at closest approach, absolute magnitudes and phase slopes constrain size and
albedo for exposure planning, and \como{} colours provide a first-order
taxonomic prior for sequencing spectroscopy. The rendezvous target \justitia{}
is already well characterised; among the flybys,
\rockox{}, \moza{} and \chimaera{} are best constrained, while \westerwald{},
\ghaf{} and especially \ousha{} would benefit most from a dedicated
pre-encounter campaign.

%% file: sections/07_conclusions.tex
\section{Conclusions}
\label{sec:conclusions}

We have analysed approximately a decade of sparse \atlas{} photometry, obtained
through the \atlas{} forced photometry service, for all seven main-belt asteroid
targets of the Emirates Mission to the Asteroid Belt. Our principal conclusions
are:

\begin{enumerate}
  \item We derive homogeneous phase slope parameters in both ATLAS filters and
        \como{} colours for all seven targets (Table~\ref{tab:phot}).

  \item We recover rotation periods spanning $\sim$3.2 to $\sim$33~h. Five are
        secure --- \rockox{}, \moza{}, \westerwald{}, \chimaera{} and
        \justitia{} --- each agreeing with independent determinations
        \citep{durech2020,pal2020,pravec2024web,fleenor2007,marciniak2025}. We support the already published
      period for \ousha{}, whose periodogram is
        alias-affected, and the period of \ghaf{} remains unresolved
        (Table~\ref{tab:rotation}).

  \item Combining the \como{} colour with a $G_{\mathrm{avg}}$-based albedo class
        (via the phase-integral relation of \citealt{shevchenko2019}), we assign
        probable taxonomies: a clear C-type (\chimaera{}), three S-type
        candidates (\westerwald{}, \ghaf{}, \moza{}), two X-complex candidates
        (\rockox{}, plausibly M-type, and \ousha{}, possibly E-type), and the
        red L/D-type \justitia{}. Where these disagree with the mission's
        family-based expectation, they constitute testable pre-encounter
        predictions.

  \item \justitia{} is confirmed as the reddest target by a wide margin
        ($c-o = 0.48$); its decade-long, rotationally-averaged colour establishes
        that its extreme spectral slope is an intrinsic, stable surface property,
        supporting an outer-Solar-System origin.

  \item Convex inversion yields a unique shape and spin-state model for
        \rockox{} (retrograde pole) and two-pole solutions for \chimaera{},
        \moza{} and \justitia{}; \westerwald{}'s spin axis is constrained to be
        retrograde.

  \item These results demonstrate the value of long-baseline survey archives
        such as ATLAS for pre-encounter physical characterisation, and identify
        \westerwald{}, \ghaf{} and \ousha{} as the targets that would most
        benefit from additional dedicated follow-up before their flybys.
\end{enumerate}

%% file: sections/08_acknowledgements.tex
\section*{Acknowledgements}

This work has made use of data from the Asteroid Terrestrial-impact Last Alert
System (ATLAS) project. The Asteroid Terrestrial-impact Last Alert System
(ATLAS) project is primarily funded to search for near earth asteroids through
NASA grants NN12AR55G, 80NSSC18K0284, and 80NSSC18K1575; byproducts of the NEO
search include images and catalogs from the survey area. This work was
partially funded by Kepler/K2 grant J1944/80NSSC19K0112 and HST GO-15889, and
STFC grants ST/T000198/1 and ST/S006109/1. The ATLAS science products have been
made possible through the contributions of the University of Hawaii Institute
for Astronomy, the Queen's University Belfast, the Space Telescope Science
Institute, the South African Astronomical Observatory (SAAO), and The Millennium
Institute of Astrophysics (MAS), Chile. This research has made use of data and services provided by the Asteroid
Lightcurve Database (LCDB; \citealt{warner2009}) and NASA's JPL Horizons ephemeris system.
N. Erasmus acknowledges financial support from the South African National Research Foundation (NRF).
J.~\v{D}urech was supported by the Czech Science Foundation grant 23-04946S.
J. Licandro and M. R. Alarcon acknowledge financial support through grant PID2024-160618NB-C22, “Hydrated Minerals and Organic Compounds in Primitive Asteroids 2”, funded by MICIU/AEI/10.13039/501100011033 and by ERDF, EU, and through the Severo Ochoa Centre of Excellence accreditation awarded to the Instituto de Astrof\'{\i}sica de Canarias, grant CEX2025-001609-S, funded by MICIU/AEI/10.13039/501100011033.
Large language models (Claude Opus 4.8) were used for language editing but all content and citations were thoroughly reviewed and verified for accuracy by the first author.

\section*{Data Availability}

The ATLAS forced photometry underlying this work is publicly available through
the ATLAS forced photometry service\footnote{\url{https://fallingstar-data.com/forcedphot/}}.

\section*{Software}

This work made use of
\texttt{astropy} \citep{astropy2022},
\texttt{numpy} \citep{harris2020},
\texttt{scipy} \citep{virtanen2020}, and
\texttt{matplotlib} \citep{hunter2007}.

%% file: sections/A1_figures.tex
\section{Per-target diagnostic plots}
\label{app:figures}

This appendix collects the six-panel \atlas{} diagnostic plots for all seven
targets, in \mbr{} encounter order. The panel layout is described at the start
of Section~\ref{sec:results}.

\begin{figure*}
\centering
\includegraphics[width=\textwidth]{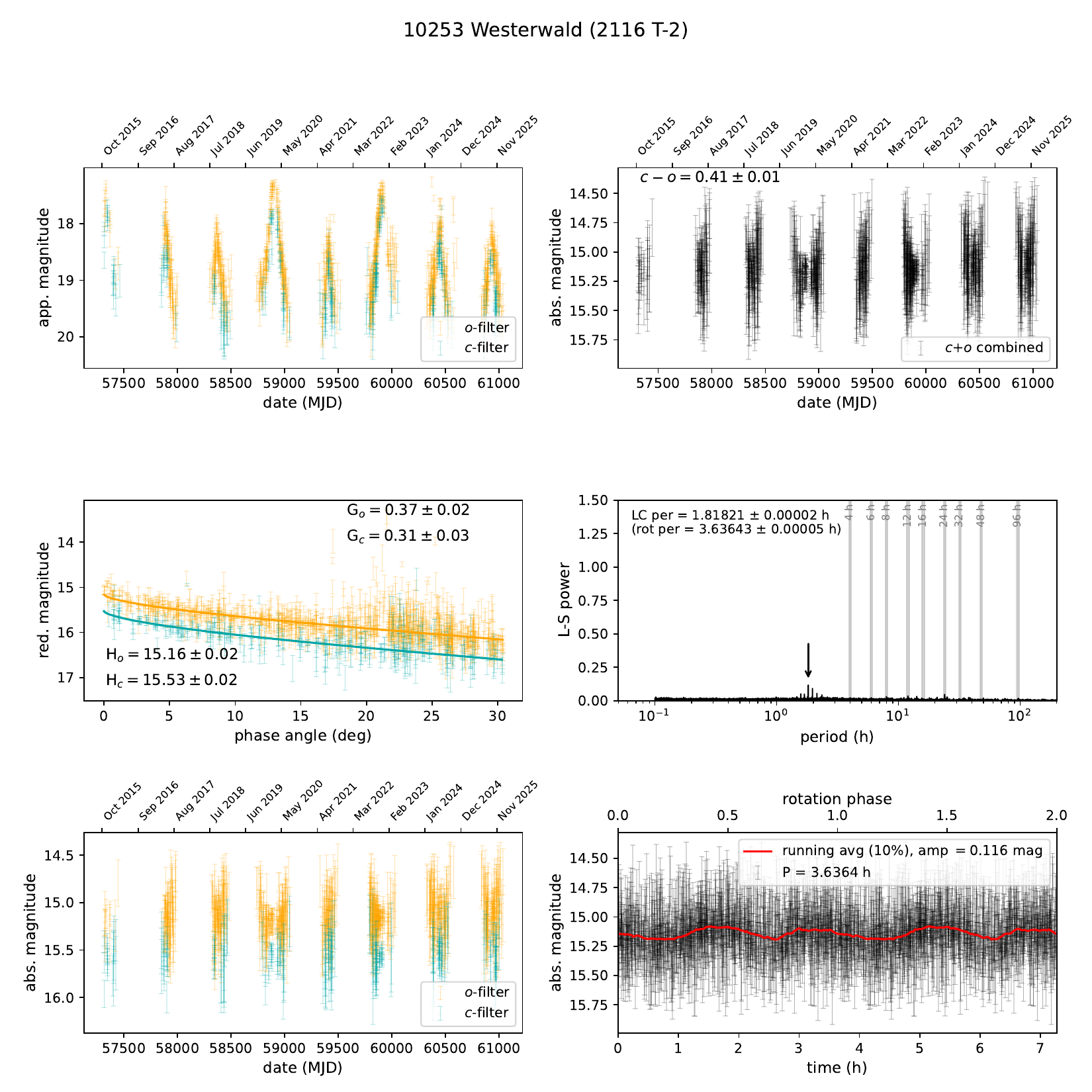}
\caption{\atlas{} photometry and derived parameters for \westerwald{}. Panels
are described at the start of Section~\ref{sec:results}.}
\label{fig:westerwald}
\end{figure*}

\begin{figure*}
\centering
\includegraphics[width=\textwidth]{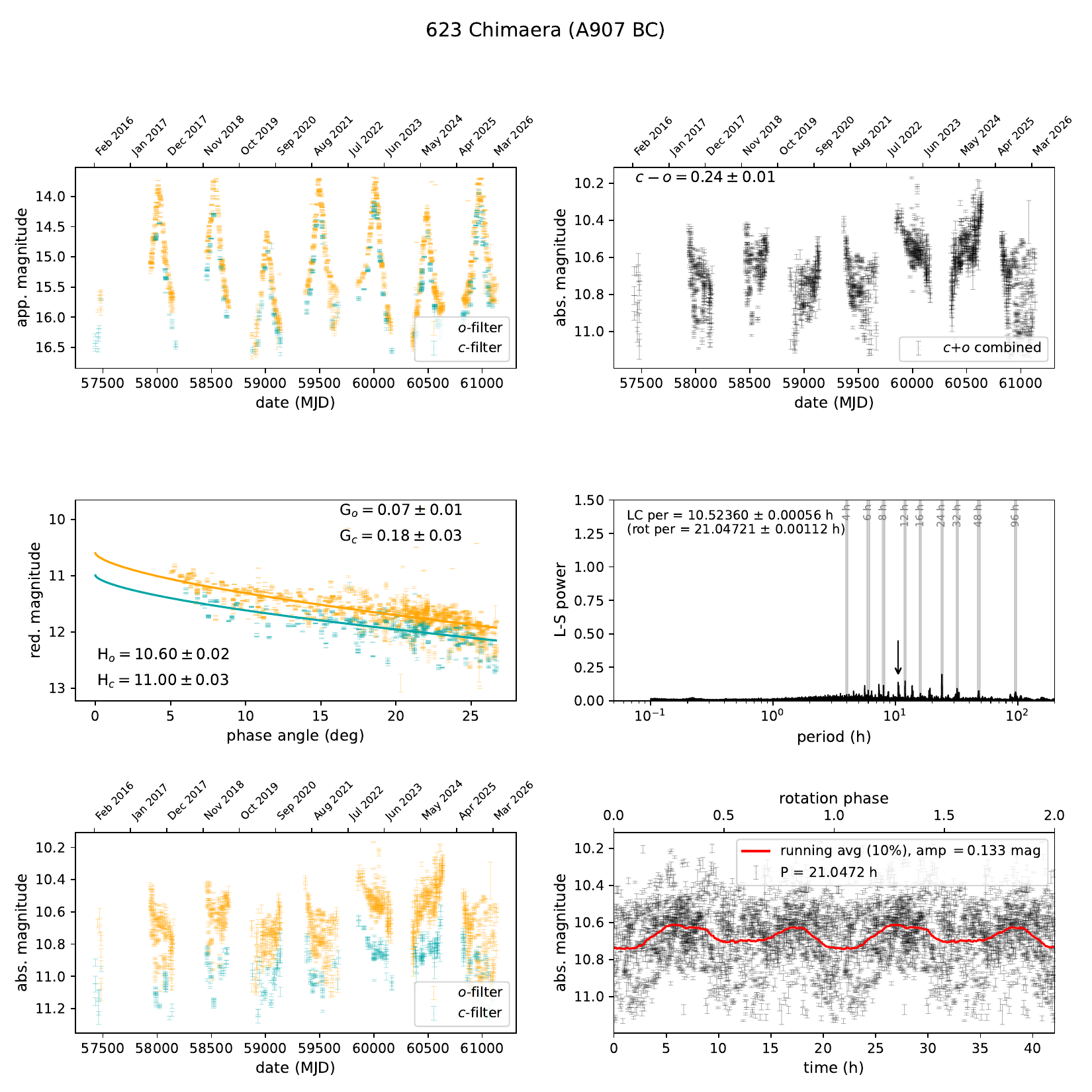}
\caption{As Figure~\ref{fig:westerwald}, for \chimaera{}.}
\label{fig:chimaera}
\end{figure*}

\begin{figure*}
\centering
\includegraphics[width=\textwidth]{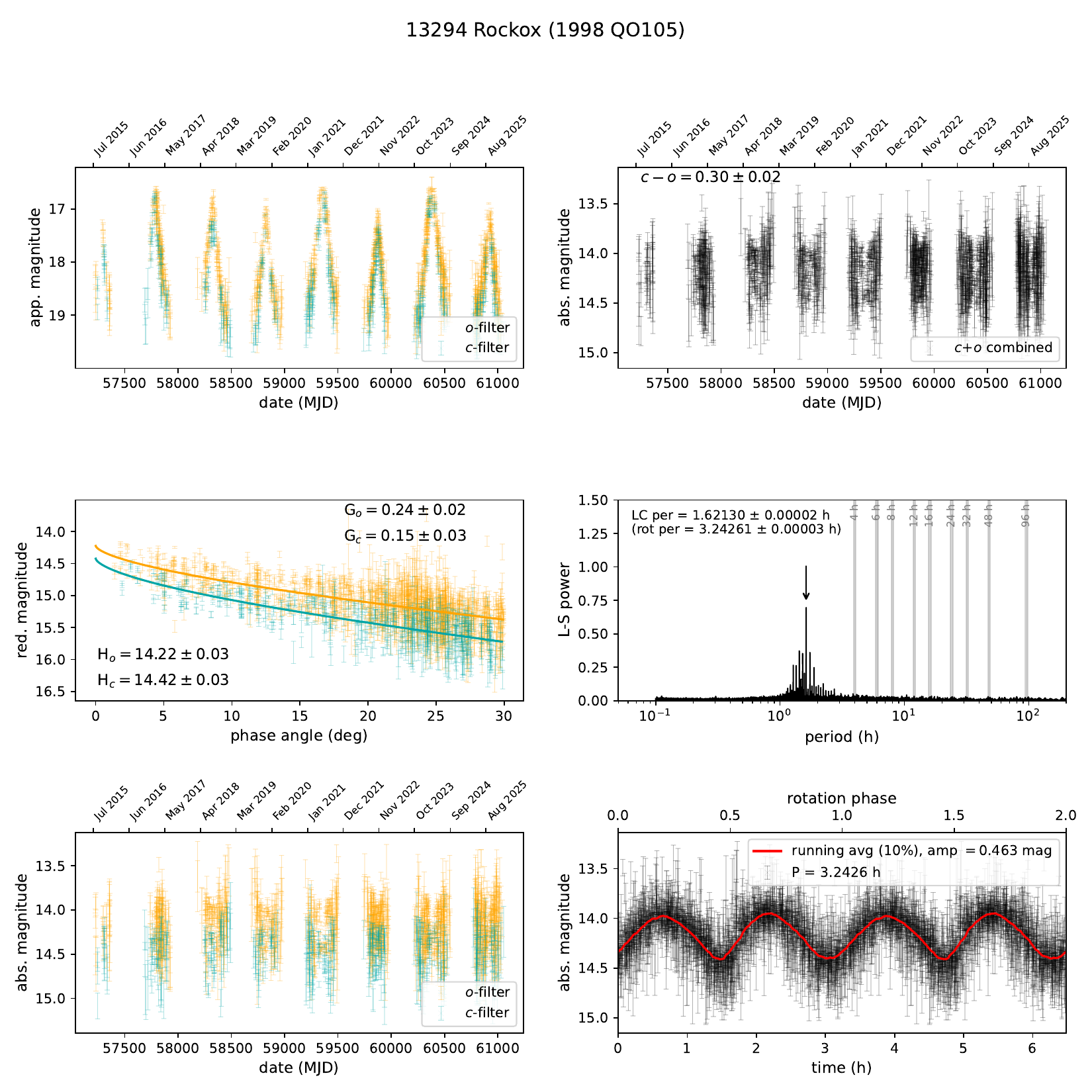}
\caption{As Figure~\ref{fig:westerwald}, for \rockox{}.}
\label{fig:rockox}
\end{figure*}

\begin{figure*}
\centering
\includegraphics[width=\textwidth]{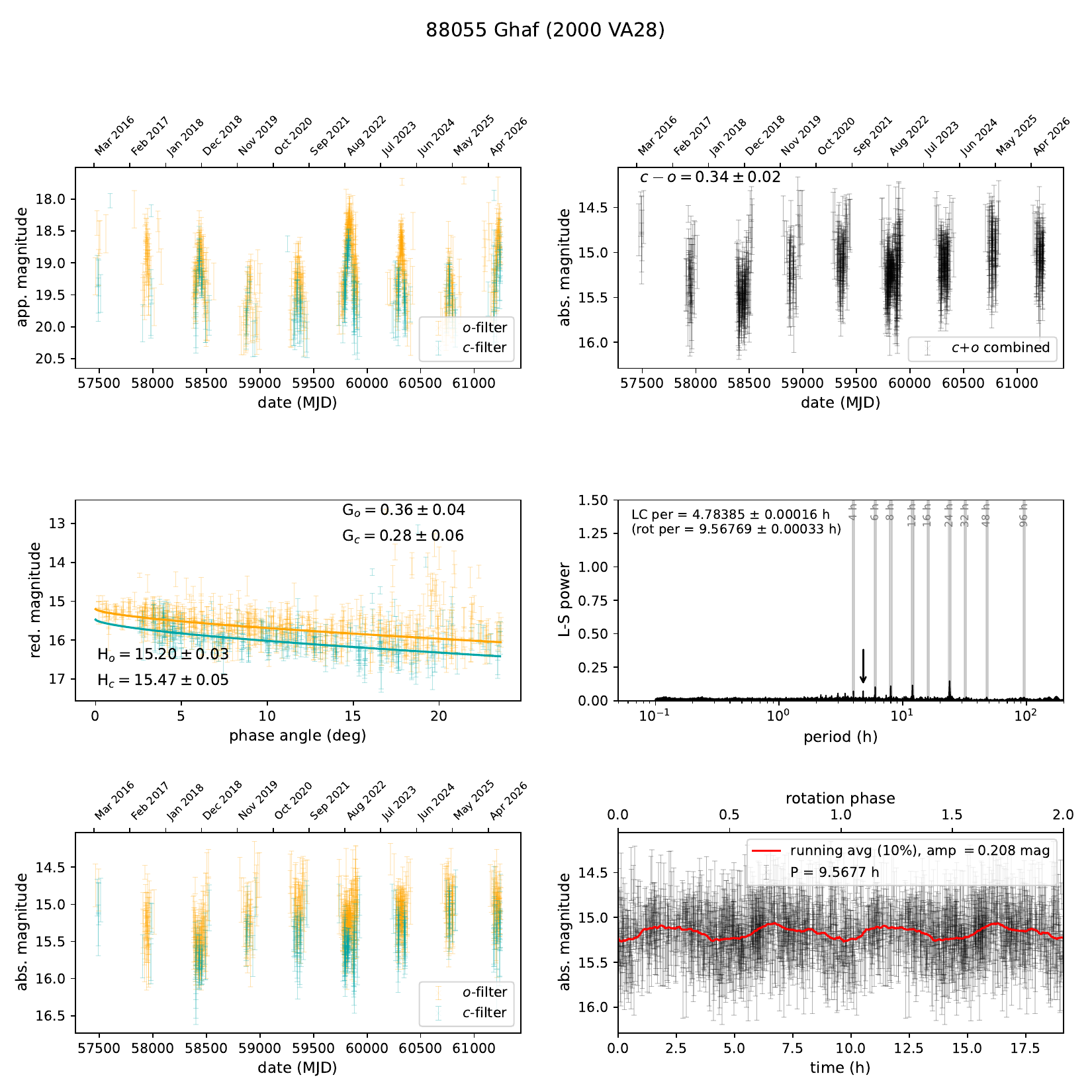}
\caption{As Figure~\ref{fig:westerwald}, for \ghaf{}.}
\label{fig:ghaf}
\end{figure*}

\begin{figure*}
\centering
\includegraphics[width=\textwidth]{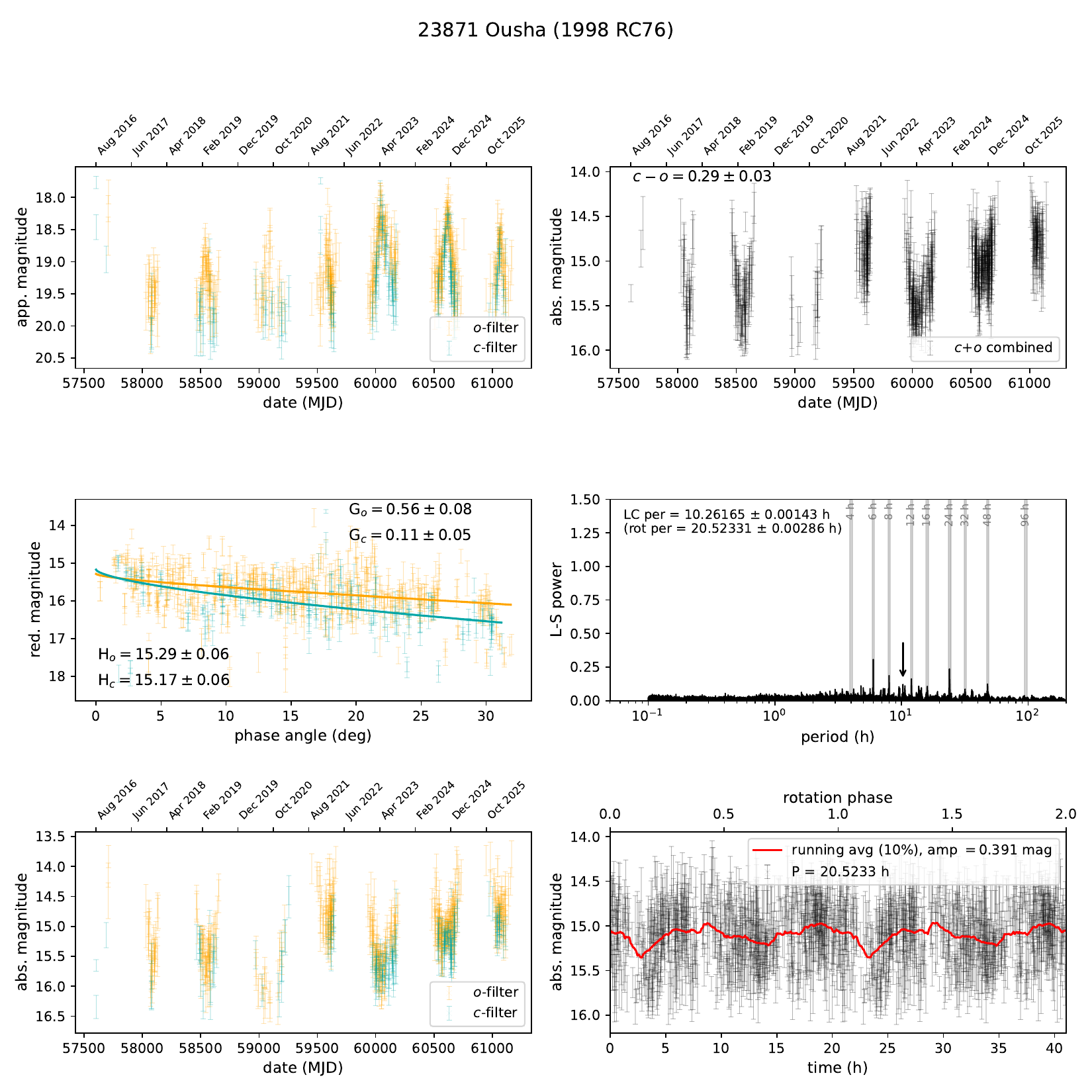}
\caption{As Figure~\ref{fig:westerwald}, for \ousha{}.}
\label{fig:ousha}
\end{figure*}

\begin{figure*}
\centering
\includegraphics[width=\textwidth]{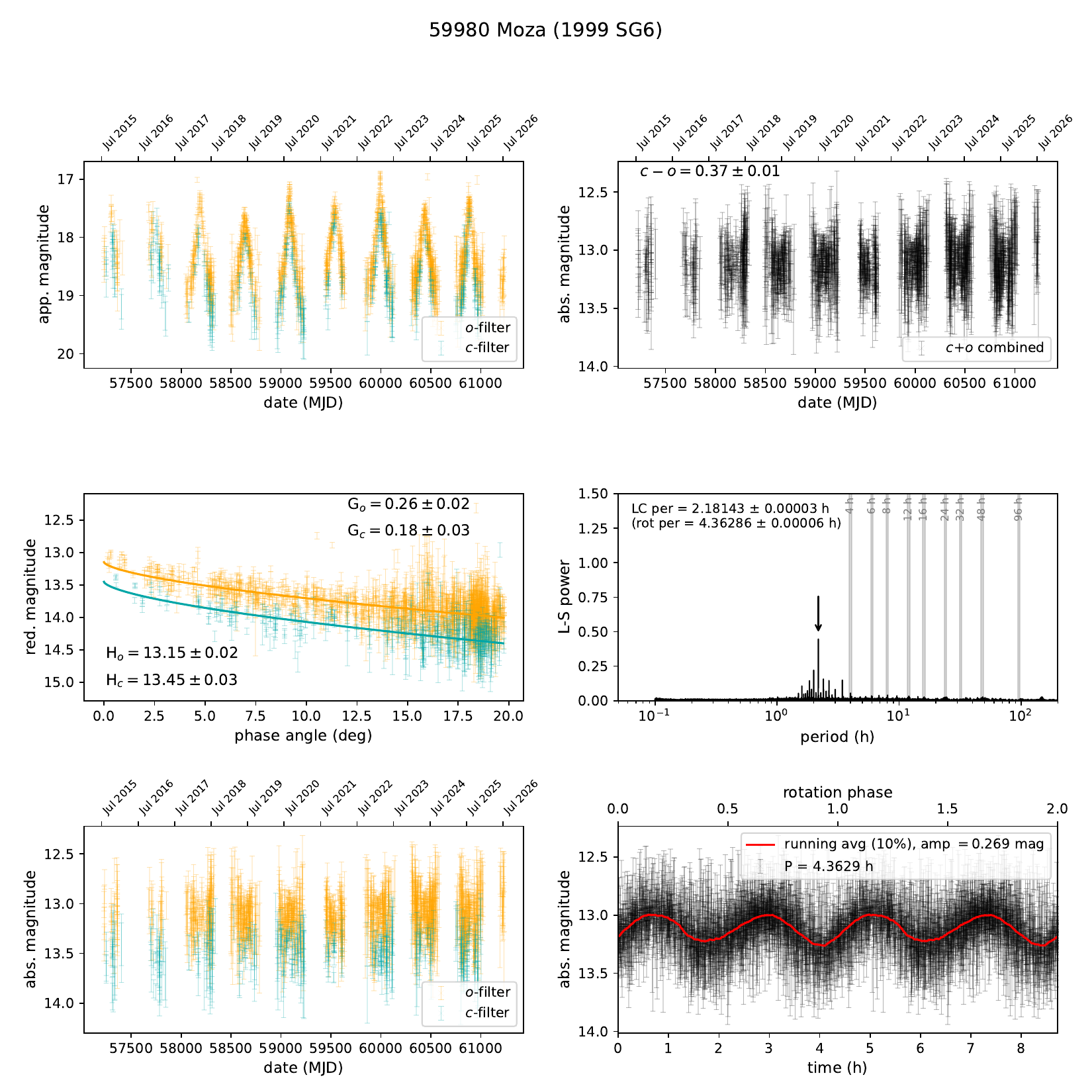}
\caption{As Figure~\ref{fig:westerwald}, for \moza{}.}
\label{fig:moza}
\end{figure*}

\begin{figure*}
\centering
\includegraphics[width=\textwidth]{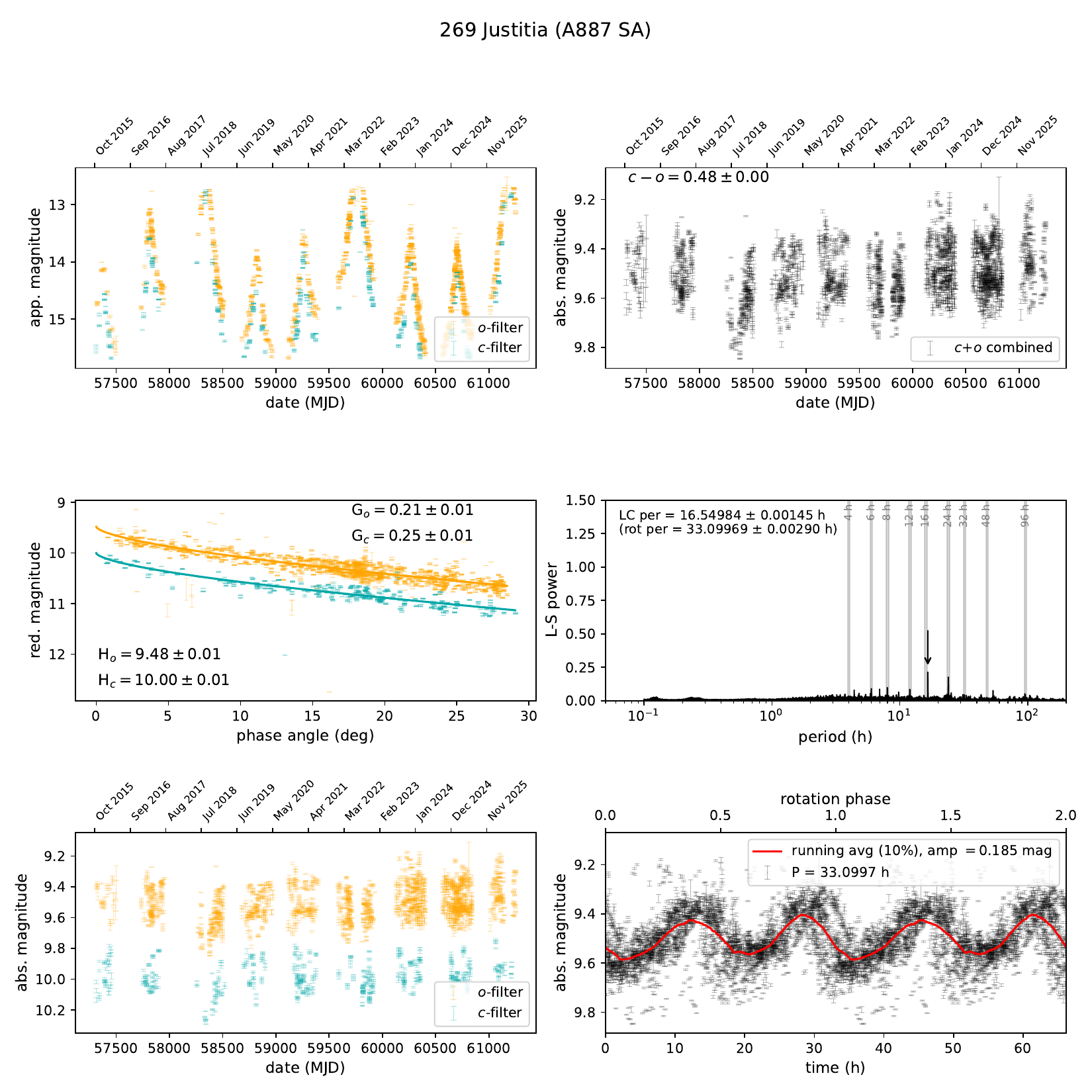}
\caption{As Figure~\ref{fig:westerwald}, for \justitia{}, the rendezvous and
landing target.}
\label{fig:justitia}
\end{figure*}